\documentclass[english,review]{elsarticle}
\usepackage{mathpazo}

\usepackage[T1]{fontenc}
\usepackage[latin9]{inputenc}
\usepackage{geometry}
\usepackage{color}
\usepackage{amsmath}
\usepackage{amssymb}
\usepackage{cancel}
\usepackage{graphicx}
\usepackage{esint}

\makeatletter

\providecommand{\tabularnewline}{\\}

\usepackage{xcolor}

\@ifundefined{showcaptionsetup}{}{%
 \PassOptionsToPackage{caption=false}{subfig}}
\usepackage{subfig}
\makeatother

\usepackage{babel}
\begin{document}

\title{An asymptotic-preserving 2D-2P relativistic Drift-Kinetic-Equation
solver for runaway electron simulations in axisymmetric tokamaks}

\author[]{L. Chacón\corref{cor1}}

\ead{chacon@lanl.gov}

\author[]{D. Daniel}

\author[]{W. T. Taitano}

\cortext[cor1]{Corresponding author}

\address{Los Alamos National Laboratory, Los Alamos, NM 87545}
\begin{abstract}
We propose an asymptotic-preserving (AP), uniformly convergent numerical
scheme for the relativistic collisional Drift-Kinetic Equation (rDKE)
to simulate runaway electrons in axisymmetric toroidal magnetic field
geometries typical of tokamak devices. The approach is derived from
an exact Green's function solution with numerical approximations of
quantifiable impact, and results in a simple, two-step operator-split
algorithm, consisting of a collisional Eulerian step, and a Lagrangian
orbit-integration step with analytically prescribed kernels. The AP
character of the approach is demonstrated by analysis of the dominant
numerical errors, as well as by numerical experiments. We demonstrate
the ability of the algorithm to provide accurate answers regardless
of plasma collisionality on a circular axisymmetric tokamak geometry.
\end{abstract}
\begin{keyword}
Relativistic drift-kinetic equation \sep Runaway electrons \sep
Relativistic collisions \sep Vlasov-Fokker-Planck \sep Asymptotic
preservation \sep Semi-Lagrangian algorithms \PACS
\end{keyword}
\maketitle

\section{Introduction}

We consider a solver designed to simulate runaway electrons produced
by a large loop voltage in axisymmetric tokamak geometry. In tokamaks,
the loop voltage induced during disruptions can produce a large amount
of runaway electrons, which may severely damage plasma facing materials
\citep{Woods,CH75}. The generated runaway current is also affected
by secondary mechanisms such as energy transfers from the primary
runaway electron current to the thermal electrons through knock-on
(large-angle) collisions. Understanding these nonlinear mechanisms
may be essential to develop either avoidance or mitigation strategies
for runaway electrons in tokamaks.

Runaway electron production is inherently a kinetic process, and requires
a relativistic kinetic treatment for fidelity, including collisions.
While collisions of fully formed relativistic runaways are infrequent,
they are critical in their seeding, and important as well when considering
mitigation strategies. In fact, there have been a significant number
of recent studies developing and using collisional kinetic models
for runaway electron simulation (see e.g\textcolor{red}{{} }\textcolor{black}{\citep{NORSE,nilsson2015kinetic,Guo17,mcdevitt2019runaway,mcdevitt2019spatial,Harvey2000,decker2016,Landreman2014,Hesslow2018}).}
Here, we adopt the relativistic drift kinetic equation (rDKE) \citep{brizard1999nonlinear},
including relativistic collisions \citep{BK87}, as the simplest high-fidelity
model capable of capturing key runaway physics of interest. We solve
the rDKE equation in axisymmetric tokamak geometries, assuming nested
magnetic flux surfaces. Recent work \citep{mcdevitt2019runaway} has
demonstrated that, even in such idealized conditions, runaway dynamics
is far from trivial, and manifests a strong dependence on plasma collisionality.

We seek a solver for the collisional rDKE that is capable of producing
the correct asymptotic limits for arbitrary plasma collisionality.
We remark that solving the rDKE equation is challenging in both weakly
and strongly collisional limits. In the weakly collisional regime,
stiff hyperbolic transport from fast electrons along field lines dominates,
making a multiscale treatment of the rDKE challenging. In the strongly
collisional regime, fast collisional timescales become the bottleneck,
demanding an implicit, conservative treatment. In this study, we employ
a recently proposed fully implicit, conservative, optimally scalable
relativistic collision algorithm \citep{daniel2020-irfp}, which effectively
takes care of numerical stiffness originating in the collision operator.
Therefore, the focus of this study is to enable a suitable multiscale
solver able to deal with stiff hyperbolic transport when plasma conditions
demand it.

Stiff hyperbolic transport is a remarkably hard mathematical and numerical
problem. Mathematically, its asymptotic treatment is complicated.
If we term $\epsilon$ as the ratio of advective to collisional timescales,
the problematic limit is $\epsilon\rightarrow0$, when advective transport
becomes arbitrarily fast. While a well-defined slow manifold exists
(the so-called limit solution in the asymptotic-preserving (AP) literature,
which, as shown later here and elsewhere \citep{crouseilles2013asymptotic,fedele2019asymptotic},
is comprised of solutions that are constants along hyperbolic orbits),
fast oscillating components excited either initially or during the
dynamical evolution will not decay in time (since hyperbolic transport
does not possess a diffusive character), and therefore may survive
to either pollute the simulation or grind the time marching to a halt.

It is therefore of particular interest to develop an AP formulation
for fast stiff hyperbolic transport in the rDKE that is able to condition
the initial condition to project out fast oscillatory components,
and to continue doing so throughout the temporal integration (as other
physics may dynamically inject such components as the simulation progresses).
Moreover, it is of interest that the formulation be uniformly convergent
in $\epsilon$, such that accurate numerical solutions are found both
for $\Delta t/\epsilon\gg1$ and $\Delta t/\epsilon\ll1$, and in
between. Finally, it is desirable that the formulation be of efficient
implementation, that it avoid iteration on the hyperbolic component
for efficiency, and that it be in principle extensible to arbitrary
3D magnetic field geometries (out of the scope of this study, but
subject of future work).

We are, of course, not the first ones to be interested in the stiff
hyperbolic transport limit. Nevertheless, there are relatively few
studies in the literature pertaining to the development of AP formulations
for this regime \citep{crouseilles2013asymptotic,fedele2019asymptotic}.
In the plasma physics community, the limit solution of the rDKE equation
for $\epsilon\rightarrow0$ has been long identified, and there are
many sophisticated code implementations devoted to it \citep{harvey1992cql3d,Harvey2000,decker2004luke,CQL3D}
(so-called bounce-averaged Fokker-Planck codes, with the bounce-average
operator the projector on to the slow manifold of the limit solution).
However, as should be expected, bounce-average formulations fail to
capture the correct solution for sufficiently collisional plasmas
(for which $\epsilon$ is no longer small enough) \citep{mcdevitt2019runaway},
and are unable to deal with arbitrary 3D magnetic field topologies
(as they bake-in the assumption of nested magnetic-flux surfaces).

While the studies in Refs. \citep{crouseilles2013asymptotic,fedele2019asymptotic}
represent serious steps toward the AP formulation of stiff hyperbolic
transport, they do not completely settle the problem. Both correctly
identify the limit solution and propose correct AP formulations for
stiff hyperbolic transport systems. However, one requires dimensional
augmentation \citep{crouseilles2013asymptotic} (problematic when
one is already dealing with two or three spatial dimensions plus at
least two momenta dimensions), and the other is fully implicit (demanding
iteration) and requires variable augmentation \citep{fedele2019asymptotic}.
Furthermore, it asymptotes to an extremely stiff anisotropic diffusion
limit equation in high-dimensionality, which can be quite difficult
to deal with accurately on a mesh \citep{chacon2014asymptotic}, and
can be very expensive to solve iteratively. Finally, they are both
implemented in very simple geometries, and neither of them consider
realistic collision sources.

We propose an AP (in fact, uniformly convergent in $\Delta t/\epsilon$,
as we shall show) solver for the collisional rDKE in 2D magnetic field
geometries (but in principle suitable for three-dimensional geometries
with few modifications). Our approach is based on a semi-Lagrangian
Green's function formalism, previously developed for the anisotropic
diffusive transport equation \citep{chacon2014asymptotic}, and generalized
here to stiff hyperbolic transport. Our numerical scheme is derived
directly from an exact Green's function integral solution, with well-defined
approximations, and leads to a simple, two-step algorithm: an implicit
collisional update (using here the methods proposed in Ref. \citep{daniel2020-irfp}),
and a Lagrangian orbit integration step with prescribed, analytical
kernels. The Lagrangian step demands an orbit integration per Eulerian
mesh point, of predictable cost and largely independent of the mesh
resolution, and is therefore optimal algorithmically. To prevent solution
pollution by fast components, our formulation is regularized in the
Lagrangian step by regularizing the transport Green's function (a
traveling delta function) as a traveling Gaussian with a finite diffusion
coefficient. This approximation, along with a particular functional
choice of the diffusion coefficient, lends the method a slightly diffusive
character that renders the method uniformly convergent, stable, and
self-conditioning. (We note that similar regularization strategies
are common in other stiff hyperbolic AP strategies \citep{fedele2019asymptotic},
for similar reasons.) We demonstrate numerically that the targeted
physics is quite insensitive to the choice of the diffusion coefficient,
when appropriately specified.

The rest of the manuscript is structured as follows. We introduce
the rDKE equation and its asymptotic properties in Sec. \ref{sec:rDKE}.
Section \ref{sec:semilagrangian-AP} presents the AP semi-Lagrangian
formulation, along with its stability and accuracy analysis. Numerical
implementation details are given in Sec. \ref{sec:Numerical-implementation-details}.
Numerical results demonstrating the advertised properties of the scheme
in an axisymmetric circular tokamak geometry, along with the verification
of the algorithm with published data, are presented in Sec. \ref{sec:Numerical-results}.
Finally, we conclude in Sec. \ref{sec:Conclusions}.

\section{\label{sec:rDKE}Relativistic drift-kinetic equation}

We consider the relativistic drift-kinetic equation (rDKE) given by
\citep{brizard1999nonlinear}:
\begin{equation}
\partial_{t}\bar{f}_{e}+\frac{1}{B}\left[\frac{\partial}{\partial\mathbf{x}}\cdot\left({B\frac{d\mathbf{x}}{d\tau}\bar{f}_{e}}\right)_{p_{\parallel},\mu}+\frac{\partial}{\partial p_{\parallel}}\left({B\frac{dp_{\parallel}}{d\tau}\bar{f}_{e}}\right)\right]=\sum_{\beta=1}^{N_{s}}C(f_{\beta},f_{e})-\mathbf{F}_{SR}\cdot\nabla_{\mathbf{p}}f_{e},\label{eq:rDKE}
\end{equation}
along with the guiding-center equations:
\begin{equation}
\frac{d\mathbf{x}}{d\tau}=\frac{p_{\parallel}}{\gamma m_{e}}\mathbf{b}\,\,;\,\,\frac{dp_{\parallel}}{d\tau}=-eE_{\parallel}-\frac{\mu}{\gamma}\mathbf{b}\cdot\nabla B.\label{eq:rel_orbit}
\end{equation}
In these equations, $\partial_{\mathbf{x}}|_{p_{\parallel},\mu}$
indicates the spatial derivative while keeping $p_{\parallel}$ and
the relativistic magnetic moment $\mu=p_{\perp}^{2}/2m_{e}B(\mathbf{x})$
constant, $\mathbf{x}$ is the guiding-center spatial position, $\tau$
is an intrinsic orbit time, $\bar{f}_{e}(\mathbf{x},p_{\parallel},\mu,t)$
(with the dependence on $\mu$ being parametric) is the drift-kinetic
(guiding-center) electron probability distribution function (PDF),
$f_{e}(\mathbf{x},p_{\parallel},p_{\perp},t)=\bar{f}_{e}(\mathbf{x},p_{\parallel},\mu(p_{\perp},\mathbf{x}),t)$
is the PDF representation in momentum-space $\mathbf{p}=(p_{\parallel},p_{\perp})$,
$f_{\beta}$ is the PDF of a prescribed ion species, $m_{e}$ and
$e$ are the electron rest mass and charge, respectively, $\mathbf{b}=\mathbf{B}/B$
a unit vector along the magnetic field $\mathbf{B}$, $\gamma=\sqrt{1+p^{2}/(m_{e}c)^{2}}$
is the relativistic factor, $p$ is the momentum magnitude, $E_{\parallel}$
is the electric field along magnetic field lines (here considered
given), $\mathbf{F}_{SR}$ is the electron synchrotron radiation force,
and $C$ is the collision operator given by,

\begin{equation}
C(f_{\beta},f_{e})=\partial_{\vec{{p}}}\cdot\left[\overline{{\overline{{D}}}}_{\beta}\cdot\partial_{\vec{{p}}}f_{e}-\frac{{m_{e}}}{m_{\beta}}\vec{{F_{\beta}}}f_{e}\right],\label{eq:collision}
\end{equation}
where $\overline{{\overline{{D}}}}_{\beta}$ represents the collisional
diffusion tensor coefficients and $\vec{{F}}_{\beta}$ represents
the collisional friction vector coefficients (computed based on the
appropriate background species $f_{\beta}$ with mass $m_{\beta}$).
The theoretical details of the collision operator are given in \citep{BK87},
and our implicit, conservative numerical treatment is described in
detail in Ref. \citep{daniel2020-irfp}. Though Eq. (\ref{eq:rDKE})
in principle may be used for multiple ion species and include radiation
forces, here we only consider the evolution of electrons interacting
with themselves, ions and external electric fields (i.e., $\mathbf{F}_{SR}=0$).

Note that the orbit equations in Eq. \ref{eq:rel_orbit} satisfy phase-space
incompressibility:
\[
\frac{\partial}{\partial\mathbf{x}}\cdot\left({B\frac{d\mathbf{x}}{d\tau}}\right)_{p_{\parallel},\mu}+\frac{\partial}{\partial p_{\parallel}}\left({B\frac{dp_{\parallel}}{d\tau}}\right)=0.
\]
 This allows us to rewrite Eq. \ref{eq:rDKE} as:
\begin{equation}
\partial_{t}\bar{f}_{e}+\left.\frac{d\bar{\mathbf{z}}}{d\tau}\right|_{\mu}\cdot\nabla_{z}\bar{f}_{e}=\sum_{\beta=1}^{N_{s}}C(f_{\beta},f_{e})=C_{e}(f_{e}),\label{eq:rDKE-nc}
\end{equation}
where $C(f_{e})$ is the (self-adjoint) collisional source, and $\bar{\mathbf{z}}=(\mathbf{x},p_{\parallel},\mu)$,
with,
\begin{equation}
\left.\frac{d\bar{\mathbf{z}}}{d\tau}\right|_{\mu}=\left(\frac{d\mathbf{x}}{d\tau},\frac{dp_{\parallel}}{d\tau},0\right).\label{eq:orbit_eq}
\end{equation}
It follows that one can write Eq. \ref{eq:rDKE-nc} as:
\begin{equation}
\partial_{t}\bar{f}_{e}+\left.\partial_{\tau}\bar{f}_{e}\right|_{\mu}=C_{e}(f_{e}),\label{eq:rDKE-nc-1}
\end{equation}
where $\bar{f}_{e}=\bar{f}_{e}(\bar{\mathbf{z}}(\tau;\mathbf{x},p_{\parallel},\mu),t)$,
with $\bar{\mathbf{z}}(\tau;\mathbf{x},p_{\parallel},\mu)$ found
by integrating Eq. \ref{eq:orbit_eq}.

For a subsequent numerical treatment (which will use a representation
in $(\mathbf{x},p_{\parallel},p_{\perp})$ coordinates), and to connect
with the asymptotic limit solution (to be discussed later), it is
of interest to express the orbits $\bar{\mathbf{z}}(\tau)$ in momentum
space $[\mathbf{z}=(\mathbf{x},p_{\parallel},p_{\perp})]$ defined
by the total energy $\mathcal{E}$ and magnetic moment $\mu$, with
$\mu$ defined above, and
\begin{equation}
\mathcal{E}=m_{e}c^{2}\gamma-e\Phi,\label{eq:total_energy}
\end{equation}
with $\Phi$ the electrostatic potential. The pair $(\mathcal{E},\mu)$
are conserved invariants in the collisionless rDKE equation. In this
representation, the perpendicular momentum $p_{\perp}$ evolves according
to magnetic moment conservation ($d\mu/d\tau|_{\mathcal{E},\mu}=0$)
as:
\[
\left.\frac{dp_{\perp}}{d\tau}\right|_{\mathcal{E},\mu}=\frac{p_{\perp}}{2B}\left.\frac{d\mathbf{x}}{d\tau}\right|_{\mathcal{E},\mu}\cdot\nabla B,
\]
and therefore:
\begin{equation}
\left.\frac{d\mathbf{z}}{d\tau}\right|_{\mathcal{E},\mu}=\left(\frac{p_{\parallel}}{\gamma m_{e}}\mathbf{b},-eE_{\parallel}-\frac{\mu}{\gamma}\mathbf{b}\cdot\nabla B,\frac{p_{\perp}p_{\parallel}}{2B\gamma m_{e}}\mathbf{b}\cdot\nabla B\right)_{\mathcal{E},\mu},\label{eq:orbit_eq-E-mu}
\end{equation}
which when integrated provides the orbit $\mathbf{z}(\tau;\mathbf{x},\mathcal{E},\mu)$
{[}i.e., the orbit $\mathbf{z}=(\mathbf{x},p_{\parallel},p_{\perp})$
starting at the spatial point $\mathbf{x}$ with conserved invariants
$(\mathcal{E},\mu)${]}. Then, Eq. \ref{eq:rDKE-nc-1} can be simply
rewritten as:
\begin{equation}
\partial_{t}f_{e}+\left.\partial_{\tau}f_{e}\right|_{\mathcal{E},\mu}=C_{e}(f_{e}),\label{eq:rDKE-nc-E-mu}
\end{equation}
where $f_{e}$ is defined as:
\[
f_{e}(\mathbf{x},p_{\parallel},p_{\perp},t)=f_{e}(\mathbf{z}(\tau=0;\mathbf{x},\mathcal{E},\mu),t),
\]
with $\mathbf{z}(\tau;\mathbf{x},\mathcal{E},\mu)$ found by integrating
Eq. \ref{eq:orbit_eq-E-mu}.

It is important to note that one can also consider an alternate definition
of the orbits in terms of $(\gamma,\mu)$, $\mathbf{z}(\tau^{*};\mathbf{x},\gamma,\mu)$
(with $\tau^{*}$ the orbit pseudotime when specifying $\gamma$ and
$\mu$), by explicitly excluding the parallel electric field $E_{\parallel}$
from the $p_{\parallel}$ update. There results:
\begin{equation}
\partial_{t}f_{e}+\left.\partial_{\tau^{*}}f_{e}\right|_{\gamma,\mu}=C_{e}(f_{e})+\frac{e}{m_{e}}E_{\parallel}\partial_{p_{\parallel}}f_{e},\label{eq:rDKE-nc-gamma-mu}
\end{equation}
with:
\begin{equation}
\frac{d\mathbf{z}}{d\tau^{*}}=\left(\frac{p_{\parallel}}{\gamma m_{e}}\mathbf{b},-\frac{\mu}{\gamma}\mathbf{b}\cdot\nabla B,\frac{p_{\perp}p_{\parallel}}{2B\gamma m_{e}}\mathbf{b}\cdot\nabla B\right)_{\gamma,\mu},\label{eq:orbit_eq-gamma-mu}
\end{equation}
and:
\[
f_{e}(\mathbf{x},p_{\parallel},p_{\perp},t)=f_{e}(\mathbf{z}(\tau^{*}=0;\mathbf{x},\gamma,\mu),t).
\]
The orbits found from Eq. \ref{eq:orbit_eq-gamma-mu} exactly conserve
$\gamma$ and $\mu$. In this study, we will consider both formulations
of the rDKE, Eqs. \ref{eq:rDKE-nc-E-mu} and \ref{eq:rDKE-nc-gamma-mu}.
When treating the electric field on the Eulerian momentum mesh, the
$(\gamma,\mu)$ formulation in Eq. \ref{eq:rDKE-nc-gamma-mu} captures
better the interaction between collisions and the electric-field acceleration,
in principle allowing for larger simulation time steps without sacrificing
accuracy (this will be shown numerically in Sec. \ref{sec:Numerical-results}).
Equation \ref{eq:rDKE-nc-gamma-mu} is also more convenient for a
dimensional analysis of the rDKE, as discussed next. From now on,
we drop the subscript ``e'' in the PDF and the electron mass.

\subsection{Dimensional analysis}

We normalize Eq. \ref{eq:rDKE-nc-gamma-mu} to the thermal speed $v_{th}=\sqrt{T_{e}/m_{e}}$
(with $T_{e}$ the electron temperature), the thermal collision time
$\tau_{c}^{th}=\frac{{4\pi\epsilon_{0}^{2}m_{e}^{2}v_{th}^{3}}}{e^{4}n_{e}\ln\Lambda}$
(with $\epsilon_{0}$ the vacuum permittivity, $n_{e}$ the electron
density, and $\ln\Lambda$ the Coulomb logarithm), and an arbitrary
length scale $L$, to find:
\begin{equation}
\partial_{\hat{t}}f+\frac{1}{\epsilon}\partial_{\hat{\tau}^{*}}f=\hat{C}(f)+\frac{e}{m}\frac{\tau_{c}^{th}}{v_{th}}E_{\parallel}\partial_{\hat{p}_{\parallel}}f.\label{eq:vfp-norm-thermal}
\end{equation}
Here, $\hat{t}=t/\tau_{c}^{th}$, $\hat{\tau}^{*}=\tau^{*}/\tau_{tr}^{th}$,
$\hat{v}=v/v_{th}$, and $\epsilon=\tau_{tr}^{th}/\tau_{c}^{th}=1/\hat{\lambda}_{mfp}^{th}\ll1$
(with $\tau_{tr}^{th}=L/v_{th}$ the characteristic transit time,
and $\hat{\lambda}_{mfp}^{th}$ the normalized thermal collisional
mean free path). Also, $\hat{C}(f)$ is the normalized collision operator.
Noting that the Dreicer field is defined as $E_{D}=\frac{m}{e}\frac{v_{th}}{\tau_{c}^{th}}$
\citep{dreicer1959electron}, we find:
\begin{equation}
\partial_{\hat{t}}f+\frac{1}{\epsilon}\partial_{\hat{\tau}^{*}}f=\hat{C}(f)+\hat{E}_{\parallel}\partial_{\hat{v}_{\parallel}}f,\label{eq:vfp-norm-th-final}
\end{equation}
with $\hat{E}_{\parallel}=\frac{E_{\parallel}}{E_{D}}.$ This is the
form of the rDKE to analyze if the electric field is kept fixed in
Dreicer units, and indicates that for a given $\hat{E}$ the asymptotic
properties of the model depend solely on $\epsilon$ (which is the
ratio of the thermal transit time and thermal collision time, and
could be smaller or larger than unity, depending on plasma conditions).

However, thermal units are not the most convenient for a numerical
implementation, for which relativistic units (speed of light $c$
and the relativistic collision time $\tau_{c}^{rel}$) are preferred.
In relativistic units, Eq. \ref{eq:vfp-norm-thermal} reads:
\begin{equation}
\partial_{\tilde{t}}f+\frac{1}{\epsilon_{rel}}\partial_{\tilde{\tau}^{*}}f=\frac{\tau_{c}^{rel}}{\tau_{c}^{th}}\hat{C}(f)+\tilde{E}_{\parallel}\partial_{\tilde{p}_{\parallel}}f.\label{eq:vfp-norm-rel}
\end{equation}
Here, $\tilde{p}=p/mc$, $\tilde{t}=t/\tau_{c}^{rel}$ with $\tau_{c}^{rel}=\frac{{4\pi\epsilon_{0}^{2}m_{e}^{2}c^{3}}}{e^{4}n_{e}\ln\Lambda}$
the relativistic collision time, $\tilde{\tau}^{*}=\tau^{*}/\tau_{tr}^{rel}$
with $\tau_{tr}^{rel}=L/c$ the relativistic transit time, $\epsilon_{rel}=\tau_{tr}^{rel}/\tau_{c}^{rel}\ll1$,
and $\tilde{E}_{\parallel}=E_{\parallel}/E_{CH}$, with $E_{CH}=\frac{mc}{e\tau_{c}^{rel}}$
the Connor-Hastie critical electric field \citep{CH75}, which is
related to the Dreicer field as $E_{CH}=E_{D}(v_{th}/c)^{2}$. Also,
we note that:
\[
\frac{\tau_{c}^{rel}}{\tau_{c}^{th}}=\left(\frac{c}{v_{th}}\right)^{3}\,\,;\,\,\epsilon=\frac{\tau_{tr}^{th}}{\tau_{c}^{th}}=\frac{L}{v_{th}\tau_{c}^{th}}=\frac{c}{v_{th}}\frac{\tau_{c}^{rel}}{\tau_{c}^{th}}\frac{\tau_{tr}^{rel}}{\tau_{c}^{rel}}=\left(\frac{c}{v_{th}}\right)^{4}\epsilon_{rel}.
\]
Equation \ref{eq:vfp-norm-rel} is less transparent asymptotically
(as $\epsilon_{rel}$ is always very small, but $\epsilon$ may be
large, depending on plasma conditions), but is the correct form to
consider if the electric field is kept fixed in Connor-Hastie units.
Note that Eq. \ref{eq:vfp-norm-rel} depends only on $\epsilon_{rel}$,
$v_{th}/c$, and $\tilde{E}_{\parallel}$, and these are the inputs
to our numerical code. The normalized orbit equations in relativistic
units read:
\begin{equation}
\frac{d\tilde{\mathbf{R}}}{d\tilde{\tau}^{*}}=\frac{\tilde{p}_{\parallel}}{\gamma}\mathbf{b}\,\,;\,\,\frac{d\tilde{p}_{\parallel}}{d\tilde{\tau}^{*}}=-\frac{\tilde{p}_{\perp}^{2}}{2\gamma B}\mathbf{b}\cdot\nabla B\,\,;\,\,\frac{d\tilde{p}_{\perp}}{d\tilde{\tau}^{*}}=\frac{\tilde{p}_{\perp}\tilde{p}_{\parallel}}{2\gamma B}\mathbf{b}\cdot\hat{\nabla}B,\label{eq:rel_orbit-normalized-no-E||}
\end{equation}
with $\gamma=\sqrt{1+\tilde{p}^{2}}$. It is straightforward to show
that these orbit integrals exactly conserve $\gamma$ and $\mu$.

We finish this section by noting that, in the previous developments,
the electric field term disappears from Eq. \ref{eq:vfp-norm-th-final}
if one considers the orbit invariants, $\mathcal{E}$ and $\mu$,
and reads:
\begin{equation}
\partial_{\hat{t}}f+\frac{1}{\epsilon}\partial_{\hat{\tau}}f=\hat{C}(f),\label{eq:vfp-norm-th-final-1}
\end{equation}
with the corresponding normalized orbit equations:
\begin{equation}
\frac{d\tilde{\mathbf{R}}}{d\tilde{\tau}}=\frac{\tilde{p}_{\parallel}}{\gamma}\mathbf{b}\,\,;\,\,\frac{d\tilde{p}_{\parallel}}{d\tilde{\tau}}=-\tilde{E}_{\parallel}-\frac{\tilde{p}_{\perp}^{2}}{2\gamma B}\mathbf{b}\cdot\hat{\nabla}B\,\,;\,\,\frac{d\tilde{p}_{\perp}}{d\tilde{\tau}}=\frac{\tilde{p}_{\perp}\tilde{p}_{\parallel}}{2\gamma B}\mathbf{b}\cdot\hat{\nabla}B.\label{eq:rel_orbit-normalized-w-E||}
\end{equation}
In the sequel, we will alternate between Eqs. \ref{eq:vfp-norm-rel}
and \ref{eq:vfp-norm-th-final-1}, depending on the purpose.

\subsection{\label{subsec:Asymptotic-properties-of-fast-transport}Asymptotic
properties of fast transport}

We investigate the asymptotic properties of Eq. \ref{eq:vfp-norm-th-final-1}
by considering the simplest collision operator model, a BGK collision
operator \citep{bgk}:
\begin{equation}
\partial_{t}f+\frac{1}{\epsilon}\partial_{\tau}f=f_{eq}-f,\label{eq:BGK-model}
\end{equation}
where we have dropped the hats, and $f_{eq}$ is some equilibrium
PDF of interest (typically, a Maxwellian distribution function). Asymptotically,
we are interested in a \emph{quiescent} (i.e., slowly evolving) solution
of the form:
\begin{equation}
f=f_{0}+\epsilon f_{1},\label{eq:asympt_exp}
\end{equation}
with $\partial_{t}f_{0}$ and $f_{1}\sim\mathcal{O}(1)$. There results:
\begin{equation}
\partial_{\tau}f_{0}\sim\mathcal{O}(\epsilon),\label{eq:0th-order-asympt}
\end{equation}
i.e., $f_{0}$ is a constant along orbits (which is the kernel of
the fast advection operator). The evolution equation for $f_{0}$
is found by integrating Eq. \ref{eq:BGK-model} along the orbit $\left\langle \cdots\right\rangle =\frac{1}{L}\int_{0}^{L}d\tau[\cdots]$
(note that this is the projection operator to the null space, and
that $\left\langle f_{0}\right\rangle =f_{0}$), to find:
\begin{equation}
\partial_{t}f_{0}=\left\langle f_{eq}\right\rangle -f_{0},\label{eq:BGK_limit_solution}
\end{equation}
which is the so-called asymptotic limit equation, which for this simple
case can be integrated exactly to give:
\[
f_{0}(t)=\left\langle f_{eq}\right\rangle -\left(\left\langle f_{eq}\right\rangle -\left\langle f(t=0)\right\rangle \right)e^{-t}.
\]
Off null-space solution components to Eq. \ref{eq:BGK-model} can
be found by a Fourier analysis, and read for the Fourier mode $e^{ik\tau}$:
\begin{equation}
f_{k}(t)=\left(f_{k}(0)-\frac{f_{eq,k}}{ik/\epsilon+1}\right)e^{-(ik/\epsilon+1)t}+\frac{f_{eq,k}}{ik/\epsilon+1}.\label{eq:off-null-space-sol}
\end{equation}
This solution exposes the key asymptotic challenge of Eq. \ref{eq:BGK-model}
when $\epsilon\rightarrow0$, namely, the \emph{asymptotic conditioning}
of the initial condition, as noted in Ref. \citep{crouseilles2013asymptotic}.
In particular, it shows that, as $\epsilon\rightarrow0$,
\begin{equation}
f_{k}(t)\rightarrow f_{k}(0)e^{-ikt/\epsilon}+\mathcal{O}(\epsilon),\label{eq:off-null-space-limit}
\end{equation}
which may or may not be small, depending on the nature of the initial
condition, but will oscillate arbitrarily fast. In order for the asymptotic
expansion in Eq. \ref{eq:asympt_exp} to hold, we must therefore have:
\[
f_{k}(0)\sim\mathcal{O}(\epsilon).
\]
Otherwise, Eq. \ref{eq:off-null-space-limit} predicts that $f_{k}(t)$
will not be sufficiently small, spoiling the asymptotics. This is
expected, as the advection operator does not allow for solution decay,
only transport. However, numerically, it must be avoided via asymptotic
conditioning of unresolved hyperbolic timescales.

While the asymptotic conditioning of Eq. \ref{eq:BGK-model} need
only be performed at the beginning of the simulation, more general
sources like the true collisional source in Eq. \ref{eq:vfp-norm-th-final-1}
and/or other physics may inject off-null-space components \emph{during}
the temporal evolution of the solution. Therefore, a suitable numerical
scheme must have the ability to recondition the solution asymptotically
\emph{throughout} the temporal integration. This, in turn, motivates
some of the numerical strategies discussed later in this study.

\subsection{Asymptotic limit equation for the rDKE: bounce-averaging}

We consider next the generalization of the limit equation and corresponding
projector operator of the simple model in the previous section, Eq.
\ref{eq:BGK_limit_solution}, to the rDKE. We begin with the non-relativistic
collisional drift-kinetic equation (DKE) in the Eulerian Frame:
\begin{equation}
\partial_{t}\bar{f}+v_{\parallel}\mathbf{b}\cdot\nabla\bar{f}|_{p_{\parallel},\mu}-\mu\mathbf{b}\cdot\nabla B\partial_{p_{\parallel}}\bar{f}-eE_{\parallel}\partial_{p_{\parallel}}\bar{f}=C(f).\label{eq:DKE}
\end{equation}
Here, $E_{\parallel}=-\mathbf{b}\cdot\nabla\Phi$, with $\Phi$ the
electrostatic potential, and $p_{\parallel}=mv_{\parallel}$. We consider
the transformation of the DKE to the kinetic-energy/magnetic-moment
space. In the drift-kinetic limit, particle orbits have two invariants,
magnetic moment $\mu=\frac{p_{\perp}^{2}}{2Bm}$ and total energy
$\mathcal{E}=E_{k}-e\Phi$, $E_{k}=p^{2}/2m$ (with $p=mv$). It follows
that:
\begin{equation}
p_{\parallel}^{2}=2m\left(\mathcal{E}+e\Phi(\mathbf{x})-\mu B(\mathbf{x})\right),\label{eq:v_par_def}
\end{equation}
which is only a function of the orbit invariants and the local magnetic
field:
\begin{equation}
\bar{f}(\mathbf{x},p_{\parallel},\mu,t)=\bar{f}(\mathbf{x},p_{\parallel}(\mathcal{E},\mu,\Phi(\mathbf{x}),B(\mathbf{x})),\mu,t)=\tilde{f}(\mathbf{x},\mathcal{E},\mu).\label{eq:f-xform}
\end{equation}
Consider the chain rule:
\[
\left.\nabla\tilde{f}\right|_{\mathcal{E},\mu}=\left.\nabla\bar{f}\right|_{p_{\parallel,\mu}}+\left.\partial_{p_{\parallel}}\bar{f}\right|_{\mathbf{x}}\nabla p_{\parallel}.
\]
Here, from Eq. \ref{eq:v_par_def} and the definition of the magnetic
moment, we have:
\[
\nabla p_{\parallel}=\frac{-\mu\nabla B+e\nabla\Phi}{v_{\parallel}}.
\]
It follows that:
\[
v_{\parallel}\mathbf{b}\cdot\nabla\bar{f}|_{p_{\parallel},\mu}=v_{\parallel}\mathbf{b}\cdot\left.\nabla\tilde{f}\right|_{\mathcal{E},\mu}+\left.\partial_{p_{\parallel}}\bar{f}\right|_{\mathbf{x}}\left(\mu\mathbf{b}\cdot\nabla B+eE_{\parallel}\right).
\]
Therefore, the transformed Vlasov equation in $(\mathcal{E},\mu)$
coordinates reads:
\begin{equation}
\partial_{t}\bar{f}+v_{\parallel}\mathbf{b}\cdot\nabla\bar{f}|_{p_{\parallel},\mu}-\mu\mathbf{b}\cdot\nabla B\partial_{p_{\parallel}}\bar{f}-eE_{\parallel}\partial_{p_{\parallel}}\bar{f}=\partial_{t}\tilde{f}+v_{\parallel}\mathbf{b}\cdot\left.\nabla\tilde{f}\right|_{\mathcal{E},\mu}=C(f),\label{eq:transformed_DKE}
\end{equation}
where only the advective term along orbits survives on the left hand
side, as expected. At this point, we are ready to perform the bounce-average
procedure of the DKE. We begin by computing the Jacobian of the ($p_{\parallel},p_{\perp}$)-($\mathcal{E},\mu$)
transformation:
\[
\mathcal{J}=\det\left[\frac{\partial(\mathcal{E},\mu)}{\partial(p_{\parallel},p_{\perp})}\right]=\frac{p_{\parallel}p_{\perp}}{m^{2}B(\mathbf{x})}.
\]
Therefore, according to the transformation of probabilities:
\begin{equation}
fp_{\perp}dp_{\perp}dp_{\parallel}=gd\mathcal{E}d\mu\Rightarrow g(\mathcal{E},\mu)=\frac{m^{2}B(\mathbf{x})}{p_{\parallel}}f(\mathbf{x},p_{\parallel},p_{\perp}).\label{eq:ba_g}
\end{equation}
Along a classical ($\mathcal{E},\mu$) orbit, $ds=\frac{p_{\parallel}}{m}d\tau$,
with $s$ the arc-length and $\tau$ (as before) the intrinsic orbit
time. We conclude:
\[
g\left.\frac{ds}{B}\right|_{\mathcal{E},\mu}=mf\left.d\tau\right|_{\mathcal{E},\mu}\Rightarrow g(\mathcal{E},\mu)=m\frac{\int_{\mathcal{E},\mu}fd\tau}{\int_{\mathcal{E},\mu}\frac{ds}{B}},
\]
where the integrals are closed for closed orbits (e.g., for nested
flux surfaces with appropriate parametrization, as is typically considered
in bounce-averaged codes), and open (from $-\infty$ to $\infty$)
otherwise. This result motivates the definition of the bounce-averaging
operator as:
\begin{equation}
\left\langle \cdots\right\rangle _{\mathcal{E},\mu}=m\frac{\int_{\mathcal{E},\mu}d\tau[\cdots]}{\int_{\mathcal{E},\mu}ds/B}.\label{eq:ba_op-E_mu}
\end{equation}
Applying the bounce-average operator (Eq. \ref{eq:ba_op-E_mu}) to
the transformed DKE, Eq. \ref{eq:transformed_DKE}, finally gives
the asymptotic limit equation sought:
\begin{equation}
\partial_{t}g(\mathcal{E},\mu)=\left\langle C(f)\right\rangle _{\mathcal{E},\mu}.\label{eq:dke_limit_solution}
\end{equation}
If the electric field is not included in the orbit definition, then
the limit equation reads:
\begin{equation}
\partial_{t}g(E_{k},\mu)=\left\langle C(f)+E_{\parallel}\partial_{p_{\parallel}}f\right\rangle _{E_{k},\mu}.\label{eq:dke_limit_solution-1}
\end{equation}
The derivation of the bounce-averaging operator for relativistic
orbits ($\mathcal{E}=mc^{2}\gamma-e\Phi,\mu$) {[}or ($mc^{2}\gamma,\mu$){]}
is in \ref{sec:Bounce-averaging-of-rDKE}, and yields the same operator
as in Eq. \ref{eq:ba_op-E_mu}, but with $ds=\frac{p_{\parallel}}{\gamma m}d\tau$.

We propose next a semi-Lagrangian operator-split algorithm for the
rDKE equation that features the orbit-averaging operator, Eq. \ref{eq:ba_op-E_mu},
as the projector to the slow manifold, thereby ensuring that the approach
be asymptotic-preserving. Moreover, we will show that the approach
is in fact uniformly convergent for arbitrary $\Delta t/\epsilon$.

\section{\label{sec:semilagrangian-AP}Asymptotic-preserving Green's function
based semi-Lagrangian formulation}

We consider the generic phase-space advection problem in $\mathbf{z}=(\mathbf{x},p_{\parallel},p_{\perp})$
with a source of the form:
\begin{equation}
\partial_{t}f+\frac{1}{\epsilon}\partial_{\tau}f=S(\mathbf{z}(\tau),t),\label{eq:vfp-model}
\end{equation}
with $\mathbf{z}(\tau)$ found by integration from either Eq. \ref{eq:orbit_eq-E-mu}
or \ref{eq:orbit_eq-gamma-mu}. This equation admits the Green's function:
\[
G\left(\tau,\frac{t}{\epsilon}\right)=\delta\left(\tau-\frac{t}{\epsilon}\right),
\]
and the \emph{exact} solution for open orbits (for closed orbits,
the integrals in $\tau'$ are closed):
\begin{equation}
f(\mathbf{z},t)=\int_{-\infty}^{\infty}d\tau'f[t=0,\hat{\mathbf{z}}(\tau';\mathbf{z})]G\left(-\tau',\frac{t}{\epsilon}\right)+\int_{-\infty}^{\infty}d\tau'\int_{0}^{t}dt'S(\hat{\mathbf{z}}(\tau';\mathbf{z}),t')G\left(-\tau',\frac{t-t}{\epsilon}'\right).\label{eq:lag-solution}
\end{equation}
Here, $\hat{\mathbf{z}}(\tau';\mathbf{z})$ is the orbit passing through
$\mathbf{z}$ at $\tau'=0$ (i.e., $\hat{\mathbf{z}}(0;\mathbf{z})=\mathbf{z}$).
Proof that Eq. \ref{eq:lag-solution} is an exact solution of Eq.
\ref{eq:vfp-model} is straightforward by direct substitution.

However, the exact solution in Eq. \ref{eq:lag-solution} does not
yet lead to a practical numerical algorithm, owing to the double integral
in the source term, and to the singular nature of the Green's function
(which leads to exact transport of all Fourier modes, as described
earlier). A practical algorithm necessitates two approximations: 1)
the regularization of the Green's function to allow for the asymptotic
conditioning of the numerical solution, and 2) a discrete formulation
conducive to a practical algorithm. We discuss these next.

\subsection{Green's function regularization}

Because we are interested in quiescent asymptotics, we relax the (singular)
delta-function Green's function into a propagating Gaussian of the
form:
\[
G\left(\tau,\xi,D\right)=\frac{1}{\sqrt{4\pi\xi D}}e^{-(\tau-\xi)^{2}/4D\xi},
\]
with $\xi=t/\epsilon$, and $D$ a (to be determined) diffusion coefficient.
The delta function is recovered in the limit of $D\rightarrow0$.
This regularized Green's function solves an advection-diffusion equation
along orbits of the form:
\[
\partial_{\xi}f+\partial_{\tau}f-D\partial_{\tau}^{2}f=0.
\]
The diffusion term is responsible for the asymptotic conditioning
of the solution at the beginning of (and during) the simulation. It
is important to emphasize that diffusion occurs strictly along orbits,
and is therefore highly anisotropic, with no cross-orbit numerical
pollution. The diffusion process is completely taken care of by the
Green's function, and introduces no numerical stiffness nor additional
implementation challenges.

Choosing the correct form of the diffusion coefficient $D$ is critical
to produce an advantageous discrete scheme. The discrete formulation
is discussed next.

\subsection{Temporally discrete algorithm}

\begin{figure}
\begin{centering}
\includegraphics[width=4in]{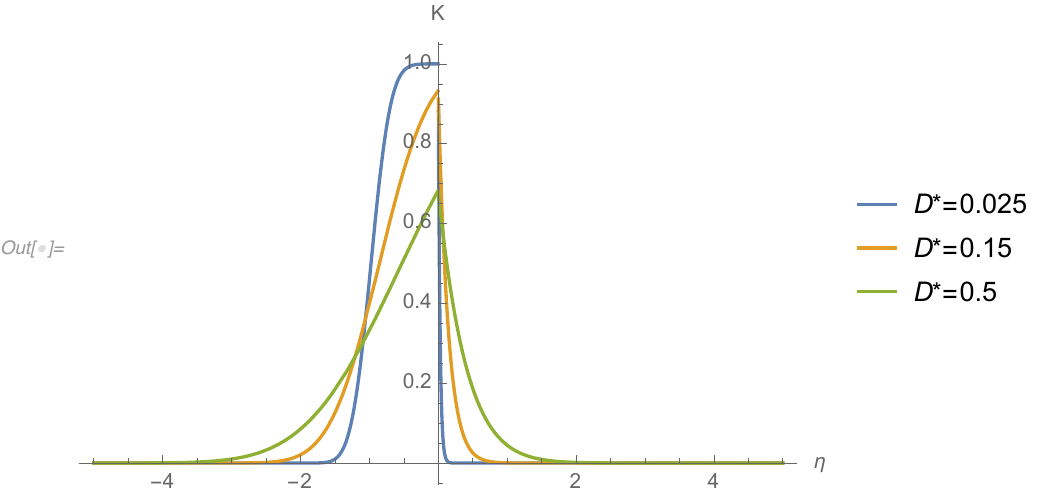}
\par\end{centering}
\caption{\label{fig:Plot-of-kernel-K}Plot of $\mathcal{K}(\eta,D^{*})$ for
several values of $D^{*}$.}
\end{figure}

The development of a discrete formulation for Eq. \ref{eq:lag-solution}
largely follows a similar development for the strongly anisotropic
heat transport equation in Ref. \citep{chacon2014asymptotic}. We
begin by reformulating Eq. \ref{eq:lag-solution} for subsequent time
steps, $t^{n}$, $t^{n+1}$, with timestep $\Delta t=t^{n+1}-t^{n}$:
\begin{equation}
f^{n+1}(\mathbf{z})=\int_{-\infty}^{\infty}d\tau'f^{n}[\hat{\mathbf{z}}(\tau';\mathbf{z})]G\left(-\tau',\frac{\Delta t}{\epsilon},D\right)+\int_{-\infty}^{\infty}d\tau'\int_{t^{n}}^{t^{n+1}}dt'S(\hat{\mathbf{z}}(\tau';\mathbf{z}),t')G\left(-\tau',\frac{t^{n+1}-t'}{\epsilon},D\right).\label{eq:lag-solution-dt}
\end{equation}
For a sufficiently small timestep $\Delta t$, we can consider the
source $S(\mathbf{z},t)$ constant, and therefore, using a $\theta$-scheme
in $\Delta t$, we can approximate \citep{chacon2014asymptotic}:
\begin{eqnarray}
\int_{-\infty}^{\infty}d\tau'\int_{t^{n}}^{t^{n+1}}dt'S(\hat{\mathbf{z}}(\tau';\mathbf{z}),t')G\left(-\tau',\frac{t^{n+1}-t'}{\epsilon},D\right) & = & \int_{-\infty}^{\infty}d\tau'S(\hat{\mathbf{z}}(\tau';\mathbf{z}),t^{n+\theta})\int_{t^{n}}^{t^{n+1}}dt'G\left(-\tau',\frac{t^{n+1}-t'}{\epsilon},D\right)\nonumber \\
 & + & \mathcal{O}\left[\Delta t^{2}\left(\frac{1}{2}-\theta+\mathcal{O}(\Delta t)\right)\right].\label{eq:constant-source}
\end{eqnarray}
The integral of the Gaussian kernel can be performed analytically,
yielding:
\begin{eqnarray*}
\mathcal{L}\left(\tau,\frac{\Delta t}{\epsilon},D\right) & = & \frac{1}{\Delta t}\int_{t^{n}}^{t^{n+1}}dt'G\left(-\tau',\frac{t^{n+1}-t'}{\epsilon},D\right)\\
 & = & \frac{\epsilon}{2\Delta t}\left[-\mathrm{sign}(\tau)-e^{-\tau/D}\left(\mathrm{{erf}}(\frac{\tau-\Delta t/\epsilon}{2\sqrt{D\Delta t/\epsilon}})-\mathrm{sign}(\tau)\right)+\mathrm{{erf}}\left(\frac{\tau+\Delta t/\epsilon}{2\sqrt{D\Delta t/\epsilon}}\right)\right].
\end{eqnarray*}
Proof is straightforward by differentiation with respect to $\Delta t/\epsilon$.
This kernel is normalized to unity, 
\[
\int_{-\infty}^{\infty}\mathcal{L}(\tau,\Delta t/\epsilon,D)d\tau=1.
\]
As we shall see in the next section, the proper form of the diffusion
coefficient is determined by numerical stability, and reads:
\begin{equation}
D=D^{*}\frac{\Delta t}{\epsilon}.\label{eq:D-choice}
\end{equation}
Defining $\eta=\epsilon\tau/\Delta t$, there results:
\[
\mathcal{L}(\tau,\Delta t/\epsilon,D^{*})=\frac{\epsilon}{\Delta t}\mathcal{K}\left(\eta,D^{*}\right),
\]
with:
\begin{equation}
\mathcal{{K}}\left(\eta,D^{*}\right)=\frac{1}{2}\left[-e^{-\eta/D^{*}}\left(\mathrm{{erf}}\left(\frac{{\eta-1}}{2\sqrt{{D^{*}}}}\right)-\mathrm{sign}(\eta)\right)\mathrm{+{erf}}\left(\frac{{\eta+1}}{2\sqrt{{D^{*}}}}\right)-\mathrm{sign}(\eta)\right].\label{eq:newI}
\end{equation}
Or, alternatively, in terms of the complementary error function (which
is better behaved numerically):
\[
\mathcal{{K}}\left(\eta,D^{*}\right)=\frac{\mathrm{sign}(\eta)}{2}\left[e^{-\eta/D^{*}}\mathrm{{erfc}}\left(\frac{{|\eta|-\mathrm{sign}(\eta)}}{2\sqrt{{D^{*}}}}\right)\mathrm{-{erfc}}\left(\frac{{|\eta|+\mathrm{sign}(\eta)}}{2\sqrt{{D^{*}}}}\right)\right].
\]
A plot of the kernel $\mathcal{K}$ is given in Fig. \ref{fig:Plot-of-kernel-K},
showing the shape dependence on $D^{*}$: it becomes box-like for
sufficiently small $D^{*}$ (asymptoting to a Heaviside function for
$D^{*}\rightarrow0$, which is the time integral of the delta function),
and spreads out as $D^{*}$ increases. 

These developments lead to the following time-discrete formulation:
\[
f^{n+1}(\mathbf{z})=\int_{-\infty}^{\infty}d\tau'f^{n}[\hat{\mathbf{z}}(\tau';\mathbf{z})]G\left(-\tau',\frac{\Delta t}{\epsilon},D^{*}\right)+\Delta t\int_{-\infty}^{\infty}d\tau'S(\hat{\mathbf{z}}(\tau';\mathbf{z}),t^{n+\theta})\mathcal{L}\left(\tau',\frac{\Delta t}{\epsilon},D^{*}\right),
\]
where:
\[
G\left(\tau,\frac{\Delta t}{\epsilon},D^{*}\right)=\frac{\epsilon}{\Delta t}\mathcal{G}(\eta,D^{*}),
\]
with:
\[
\mathcal{G}(\eta,D^{*})=\frac{1}{\sqrt{4\pi D^{*}}}e^{-(\eta-1)^{2}/4D^{*}}.
\]
Calling:
\begin{eqnarray}
\mathcal{I}\left(f^{n};\mathbf{z},\frac{\Delta t}{\epsilon},D^{*}\right) & = & \int_{-\infty}^{\infty}d\eta'f^{n}[\hat{\mathbf{z}}(\tau'=\eta'\Delta t/\epsilon;\mathbf{z})]\mathcal{G}(-\eta',D^{*}),\label{eq:I-propagator}\\
\mathcal{P}\left(S^{n+\theta};\mathbf{z},\frac{\Delta t}{\epsilon},D^{*}\right) & = & \int_{-\infty}^{\infty}d\eta'S(\hat{\mathbf{z}}(\tau'=\eta'\Delta t/\epsilon;\mathbf{z}),t^{n+\theta})\mathcal{K}\left(\eta',D^{*}\right),\label{eq:P-propagator}
\end{eqnarray}
we finally arrive to the discrete temporal update sought:
\begin{equation}
f^{n+1}(\mathbf{z})\approx\mathcal{I}\left(f^{n};\mathbf{z},\frac{\Delta t}{\epsilon},D^{*}\right)+\Delta t\mathcal{P}\left(S^{n+\theta};\mathbf{z},\frac{\Delta t}{\epsilon},D^{*}\right).\label{eq:discrete-coupled}
\end{equation}
This scheme is entirely analogous to the one proposed in Ref. \citep{chacon2014asymptotic}
for anisotropic diffusion. A generalization to second-order backward
differentiation formulas can be found in the reference.

However, Eq. \ref{eq:discrete-coupled} remains expensive, as the
source $S^{n+\theta}$ requires evaluation of the collision operator
using the new-time solution, $f^{n+1}$, and is therefore implicitly
coupled. To resolve this difficulty, we follow the reference and consider
a first-order accurate operator-split (OS) semi-Lagrangian algorithm
as follows:
\begin{itemize}
\item Implicit Eulerian step: 
\begin{equation}
f^{*}(\mathbf{z})=f^{n}(\mathbf{z})+\Delta t\,\left(\theta S[f^{*}]+(1-\theta)S[f^{n}]\right).\label{eq:OS-step1}
\end{equation}
Choosing $\theta=1$ (backward-Euler or BDF1) is consistent with the
order of the overall scheme. This step just requires the integration
of the (generally nonlinear) collisional source, perhaps with additional
terms for the electric field or radiation forces. We accomplish this
here with the algorithm proposed in Ref. \citep{daniel2020-irfp}.
\item Lagrangian transport step: 
\begin{equation}
f^{n+1}(\mathbf{z})=\mathcal{I}\left(f^{n};\mathbf{z},\frac{\Delta t}{\epsilon}\right)+\mathcal{P}\left(f^{*}-f^{n};\mathbf{z},\frac{\Delta t}{\epsilon}\right).\label{eq:OS-step2}
\end{equation}
This is a post-processing step (since all fields in the orbit integrals
are known) that requires performing an orbit integration (with either
Eq. \ref{eq:orbit_eq-E-mu} or \ref{eq:orbit_eq-gamma-mu}) per phase-space
mesh point.
\end{itemize}
As we shall see in the next section, the algorithm can be made absolutely
stable by the judicious choice of $D^{*}$, and is uniformly convergent
in $\Delta t/\epsilon$. The numerical implementation details are
very similar to those in Ref. \citep{chacon2014asymptotic}, and will
be reviewed later in this study.

\subsection{Stability analysis}

\begin{figure}
\begin{centering}
\includegraphics[height=2in]{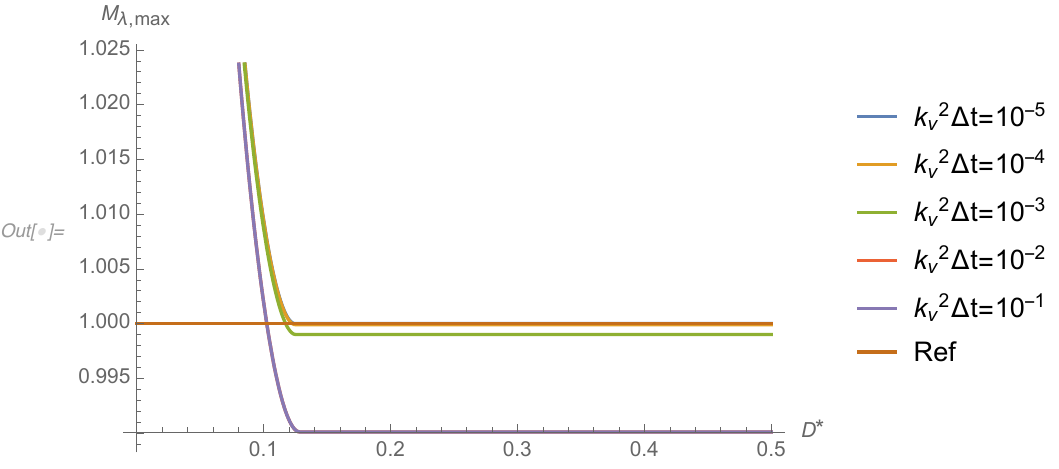}
\par\end{centering}
\caption{\label{fig:stability}Plot of maximum amplification factor vs. $D^{*}$
for various collisional strengths $S_{r}=k_{v}^{2}$.}
\end{figure}

We consider the operator-split semi-Lagrangian formulation in the
previous section using a backward Euler integration of the source
term. We assume the source is a linear functional of the dependent
variable, $f$. We assume orbits are straight in this analysis. Fourier-analyzing
Eq. \ref{eq:OS-step2} gives:
\[
f_{\mathbf{k}}^{n+1}=e^{-L(k_{\parallel}\Delta t/\epsilon,D^{*})}f_{\mathbf{k}}^{n}-\Delta t\,S_{\mathbf{k}}f_{\mathbf{k}}^{*}\frac{1-e^{-L(k_{\parallel}\Delta t/\epsilon,D^{*})}}{L(k_{\parallel}\Delta t/\epsilon,D^{*})},
\]
with $f_{\mathbf{k}}^{n+1}$ the Fourier amplitude of the new-time
PDF for wavenumber $\mathbf{k}$, $f_{\mathbf{k}}^{*}$ the same for
the solution of the first step in the OS algorithm, $k_{\parallel}=\mathbf{k}\cdot\mathbf{b}$
the wavenumber along the orbit, $S_{\mathbf{k}}=S_{r}+iS_{i}$ the
amplitude of the Fourier transform of the (linearized) source, and
\begin{equation}
L(\xi,D^{*})=i\xi+D^{*}\xi^{2}.\label{eq:L-def}
\end{equation}
The convenience of the choice of the diffusion coefficient $D$ in
Eq. \ref{eq:D-choice} is now apparent, as it leads to a very simple
functional form of the amplification factor $\frac{f_{\mathbf{k}}^{n+1}}{f_{\mathbf{k}}^{n}}$
in terms of $\xi=k_{\parallel}\Delta t/\epsilon$. In the Fourier
transform of the source, the real component corresponds to the collisional
term (which is self-adjoint), and the imaginary one to possible advective
transport terms in momentum space (e.g., E-field, radiation force).
The Fourier transform of the backward-Euler integration step (Eq.
\ref{eq:OS-step1}) gives: 
\[
f_{\mathbf{k}}^{*}=\frac{f_{\mathbf{k}}^{n}}{1+\Delta t\,S_{\mathbf{k}}}.
\]
There results the complex amplification factor:
\[
\lambda(\xi,D^{*},S_{\mathbf{k}})=\frac{f_{\mathbf{k}}^{n+1}}{f_{\mathbf{k}}^{n}}=e^{-L(\xi,D^{*})}-\frac{\Delta t\,S_{\mathbf{k}}}{1+\Delta t\,S_{\mathbf{k}}}\frac{1-e^{-L(\xi,D^{*})}}{L(\xi,D^{*})}.
\]
Stability requires:
\[
M_{\lambda}(\xi,D^{*},S_{\mathbf{k}})=|\lambda(\xi,D^{*},S_{\mathbf{k}})|<1.
\]
To find the stability boundary in $D^{*}$, we maximize $M_{\lambda}(\xi,D^{*},S_{\mathbf{k}})$
for each value of $D^{*}$, with $\Delta tS_{r}<1$ (i.e., assuming
dynamical collisional timescales are resolved, but still allowing
for implicit timesteps much larger than numerical stability constraints)
in the domain $\xi\in[0,5]$, $\Delta tS_{i}\in[0,1]$. This maximum
is found to be monotonically decreasing with $D^{*}$. The result
for $D^{*}\in[0,0.5]$ is depicted in Fig. \ref{fig:stability} for
various values of $\Delta tS_{r}=\Delta tk_{v}^{2}$ (with $k_{v}^{2}$
a measure of the magnitude of the collisional source Fourier amplitude),
and demonstrates stability for $D^{*}\gtrsim0.125$. If the source
is purely collisional, i.e., $S_{i}=0$, the same analysis indicates
stability for $D^{*}\gtrsim0.025$ (not shown). As will be shown,
the computational cost of the orbit integrals scales as $\sqrt{D^{*}}$,
and therefore the latter may be in principle more efficient. However,
as we shall see, this efficiency gain is offset by the cost of reduced
outer timesteps, required to preserve accuracy because of the implied
splitting between the electric field acceleration term and collisions.
As a result, treating acceleration terms on the momentum mesh is more
efficient in practice. 

\subsection{Asymptotic-preserving character of the semi-Lagrangian algorithm:
limit solution}

We begin with the propagator step of the operator-split semi-Lagrangian
algorithm (Eq. \ref{eq:OS-step2}):
\begin{equation}
f^{n+1}(\mathbf{z})=\mathcal{I}\left(f^{n};\mathbf{z},\frac{\Delta t}{\epsilon}\right)+\mathcal{P}\left(f^{*}-f^{n};\mathbf{z},\frac{\Delta t}{\epsilon}\right),\label{eq:semi-lag-update}
\end{equation}
where $f^{*}$ is found from updating the source only (Eq. \ref{eq:OS-step1}),
and the propagators $\mathcal{I}$ and $\mathcal{P}$ are defined
in Eqs. \ref{eq:I-propagator} and \ref{eq:P-propagator}, respectively,
and reproduced here considering an arbitrary $\eta=\epsilon\tau/\Delta t$
as the starting point of the orbit integrals (instead of $\eta=0$):
\begin{eqnarray}
\mathcal{I}\left(f^{n};\mathbf{z}(\eta),\frac{\Delta t}{\epsilon},D^{*}\right) & = & \int_{-\infty}^{\infty}d\eta'f^{n}[\hat{\mathbf{z}}(\tau'=\eta'\Delta t/\epsilon;\mathbf{z})]\mathcal{G}(\eta-\eta',D^{*}),\label{eq:I-propagator-1}\\
\mathcal{P}\left(S^{n+\theta};\mathbf{z}(\eta),\frac{\Delta t}{\epsilon},D^{*}\right) & = & \int_{-\infty}^{\infty}d\eta'S(\hat{\mathbf{z}}(\tau'=\eta'\Delta t/\epsilon;\mathbf{z}),t^{n+\theta})\mathcal{K}\left(\eta'-\eta,D^{*}\right).\label{eq:P-propagator-1}
\end{eqnarray}
The kernels $\mathcal{G}$ and $\mathcal{K}$ are normalized:
\begin{eqnarray*}
\int_{-\infty}^{\infty}d\eta\mathcal{G}(\eta-\eta',D^{*}) & = & \int_{-\infty}^{\infty}d\eta\mathcal{K}(\eta-\eta',D^{*})=1.
\end{eqnarray*}
Therefore, integrating Eqs. \ref{eq:I-propagator-1} and \ref{eq:P-propagator-1}
with respect to $\eta$, and using the definition of the bounce-averaging
operator in Eq. \ref{eq:ba_op-E_mu}, gives:
\begin{eqnarray*}
\int_{-\infty}^{\infty}d\eta\mathcal{I}\left(f^{n};\mathbf{z}(\eta),\frac{\Delta t}{\epsilon},D^{*}\right) & = & \int_{-\infty}^{\infty}d\eta'f^{n}[\hat{\mathbf{z}}(\tau'=\eta'\Delta t/\epsilon;\mathbf{z})]=\frac{\epsilon}{\Delta t}\left\langle f^{n}\right\rangle \int_{-\infty}^{\infty}ds'/B,\\
\int_{-\infty}^{\infty}d\eta\mathcal{P}(f^{*}-f^{n};\mathbf{z}(\eta),\Delta t) & = & \int_{-\infty}^{\infty}d\eta'(f^{*}[\hat{\mathbf{z}}(\tau'=\eta'\Delta t/\epsilon;\mathbf{z})]-f^{n}[\hat{\mathbf{z}}(\tau'=\eta'\Delta t/\epsilon;\mathbf{z})])=\frac{\epsilon}{\Delta t}\left\langle f^{*}-f^{n}\right\rangle \int_{-\infty}^{\infty}ds'/B.
\end{eqnarray*}
Introducing these results in Eq. \ref{eq:semi-lag-update} and using
Eq. \ref{eq:OS-step1}, yields the following first-order-accurate
discrete integral of the asymptotic limit equation:
\[
\left\langle f^{n+1}\right\rangle =\left\langle f^{*}\right\rangle =\left\langle f^{n}\right\rangle +\Delta t\left\langle S[f^{*}]\right\rangle .
\]
We have thus proved that the limit solution of the semi-Lagrangian
formulation for $\epsilon\rightarrow0$ is in fact the bounce-averaged
solution, and therefore the algorithm is asymptotic-preserving (AP).

\subsection{\label{subsec:Asymptotic-conv}Asymptotic properties of semi-Lagrangian
scheme: error analysis and uniform convergence}

We demonstrate next that the semi-Lagrangian algorithm is not only
AP, but uniformly convergent in $\Delta t/\epsilon$. We begin by
considering the rDKE equation in thermal units along relativistic
orbits ($\mathcal{E}=\gamma-\Phi,\mu$), including the regularizing
diffusion term:
\begin{equation}
\partial_{t}f+\frac{1}{\epsilon}\partial_{\tau}f-\frac{D^{*}\Delta t}{\epsilon^{2}}\partial_{\tau}^{2}f=C(f).\label{eq:vfp}
\end{equation}
As before (Sec. \ref{subsec:Asymptotic-properties-of-fast-transport}),
we begin the asymptotic analysis by considering a Hilbert expansion
for $\epsilon\ll1$:
\begin{equation}
f=f_{0}+\epsilon^{\alpha}f_{1}+\ldots\label{eq:f-asymptotic}
\end{equation}
Here, $\left\langle f\right\rangle =\left\langle f_{0}\right\rangle $
(constant along orbits), and $f_{i>0}\sim\mathcal{O}(1)$, but $\left\langle f_{i>0}\right\rangle =0$.
Introducing this expansion into Eq. \ref{eq:vfp}, we find:
\[
\partial_{t}(f_{0}+\epsilon^{\alpha}f_{1})+\epsilon^{\alpha-1}\partial_{\tau}f_{1}-\epsilon^{\alpha-2}D^{*}\Delta t\partial_{\tau}^{2}f_{1}+HOT(\epsilon)=C(f),
\]
where the source behaves regularly with $\epsilon$ (i.e., has no
$1/\epsilon$ dependence), and $\alpha\geq1$ for regularity of the
advective term (otherwise it would imply $f_{1}=0$). Taking the average
of this equation, we get the limit equation:
\[
\partial_{t}\left\langle f_{0}\right\rangle =\left\langle C(f)\right\rangle ,
\]
and therefore:
\[
\epsilon^{\alpha}\partial_{t}f_{1}+\epsilon^{\alpha-1}\partial_{\tau}f_{1}-\epsilon^{\alpha-2}D^{*}\Delta t\partial_{\tau}^{2}f_{1}+HOT(\epsilon)=C(f_{0})-\left\langle C\right\rangle =\mathcal{O}(1).
\]
The last equality is justified because $C(f_{0})=C(\left\langle f\right\rangle )\neq\left\langle C(f)\right\rangle $
in general. Since $\alpha\geq1$, matching $\epsilon$ powers, we
get:
\[
\epsilon^{\alpha-2}D^{*}\Delta t\partial_{\tau}^{2}f_{1}=\mathcal{O}(1).
\]
Since $f_{1}\sim\mathcal{O}(1)$ by \emph{ansatz}, and taking $\partial_{\tau}f_{1}=ik_{\parallel}f_{1}$,
it follows that:
\[
\epsilon^{\alpha-2}D^{*}\Delta tk_{\parallel}^{2}\sim1.
\]
This suggests $\alpha=2$ (which will be verified numerically) and
consequently that the diffusion Fourier-mode cut-off (which controls
the asymptotic transition) happens at $k_{\parallel,\mathrm{cutoff}}\sim1/(D^{*}\Delta t)^{1/2}$.
It also follows that: 
\[
\xi_{\mathrm{cutoff}}=k_{\parallel}\Delta t/\epsilon\sim\epsilon^{-1}(\Delta t/D^{*})^{1/2}\,\,;\,\,f=f_{0}+\mathcal{O}(\epsilon^{2}).
\]

With these results, we are equipped to study the temporal discretization
error of the semi-Lagrangian algorithm. Asymptotic error of convergence
is expected to be first-order, both from its OS character and the
BDF1 treatment of the Eulerian step. The sources of temporal discretization
error are the constant-source approximation and the operator splitting.
We analyze these next. The analysis exactly follows that in Ref. \citep{chacon2014asymptotic}
(Appendices B and C in the reference).

We begin with the splitting error. Ref. \citep{chacon2014asymptotic}
gives the following expression for the global truncation error (GTE,
defined as the accumulated temporal error for the total timespan of
a given simulation) due to splitting:
\[
\mathcal{E}_{split}^{GTE}\sim k_{v}^{2}\left|f_{\mathbf{k}}\right|\left|1-e^{-L(\xi,D^{*})}\right|\sim k_{v}^{2}\left|f_{\mathbf{k}}\right|\min[\xi,1].
\]
Here, $L(\xi,D^{*})$ is defined in Eq. \ref{eq:L-def}, and, as before,
$k_{v}^{2}$ is a measure of the Fourier-transform amplitude of the
(linearized and self-adjoint) collision operator. From this expression,
it is apparent that the splitting error strictly vanishes for the
limit solution ($k_{\parallel}=0$), since $\xi=0$ in that case.
For finite $k_{\parallel}$, using the results in the previous section
(and in particular that $f_{\mathbf{k}}\sim\epsilon^{2}$ for finite
$k_{\parallel}$ and small enough $\epsilon$), we readily conclude
that: 
\[
\mathcal{E}_{split}^{GTE}(k_{\parallel}\neq0)\sim\begin{cases}
\xi\sim\Delta t & \xi\ll1\\
\epsilon(\Delta t/D^{*})^{1/2} & \xi\sim\xi_{\mathrm{cutoff}}\\
\epsilon^{2} & \xi\gg1
\end{cases}.
\]
This result proves uniform convergence with respect to $\xi\sim\Delta t/\epsilon$:
it is first-order accurate (as expected) and independent of $D^{*}$
for resolved modes ($\xi\ll1$), and scales as $\epsilon^{2}$ in
the asymptotic regime for unresolved ones ($\xi\gg1$). At the transition,
we find the error scales as the geometric mean of the two, $\epsilon\sqrt{\Delta t/D^{*}}$,
showing a weak dependence on $D^{*}$. These conclusions will be verified
numerically in Sec. \ref{sec:Numerical-results}.

Since the splitting error vanishes for the limit solution ($k_{\parallel}=0$),
its error is determined by the constant-source approximation in Eq.
\ref{eq:constant-source}. Following Ref. \citep{chacon2014asymptotic},
the global truncation error for the OS+BDF1 algorithm can be written
as:
\[
\mathcal{E}_{src}^{GTE}\sim k_{v}^{2}\Delta t\left|\partial_{t}f_{\mathbf{k}}\frac{\left[1+\hat{L}(\xi)\right]e^{-\hat{L}(\xi)}-1}{[\hat{L}(\xi)]^{2}}\right|.
\]
Here,
\[
\partial_{t}f_{\mathbf{k}}\sim\frac{f_{\mathbf{k}}}{\tau_{k}},
\]
with $\tau_{k}$ the time constant for the Fourier mode $\mathbf{k}=(k_{\parallel},k_{v})$,
which is found from the original equation, Eq. \ref{eq:vfp}, as:
\[
\frac{\Delta t}{\tau_{k}}\sim|\hat{L}(\xi)+k_{v}^{2}\Delta t|.
\]
There results:
\[
\mathcal{E}_{src}^{GTE}\sim k_{v}^{2}\left|f_{\mathbf{k}}\right||\hat{L}(\xi)+k_{v}^{2}\Delta t|\left|\frac{\left[1+\hat{L}(\xi)\right]e^{-\hat{L}(\xi)}-1}{[\hat{L}(\xi)]^{2}}\right|\sim\begin{cases}
k_{v}^{4}\Delta t & \xi\ll1\\
\frac{k_{v}^{2}\left|f_{\mathbf{k}}\right|}{D^{*}\xi^{2}}\sim\epsilon^{4} & \xi\gg1
\end{cases},
\]
where we have used that $\left|f_{\mathbf{k}}\right|\sim\epsilon^{2}$
for $k_{\parallel}\neq0$ and $\xi\gg1$. The constant-source error
is first-order for resolved modes (and in particular for the limit
solution with $k_{\parallel}=0$), and scales as $\epsilon^{4}$ otherwise
(i.e., it is negligible compared to the OS error).

\section{\label{sec:Numerical-implementation-details}Numerical implementation
details}

The numerical integration of the collision operator on the Eulerian
mesh is performed as described in Ref. \citep{daniel2020-irfp}. Even
though our collisional implementation is in principle fully nonlinear,
in certain tests we have considered a linearized implementation of
the collision operator (in which the background electron population
does not evolve) for some of the comparisons below for consistency
with the original references.

The numerical implementation of the Lagrangian integrals along orbits
largely follows the details outlined in Ref. \citep{chacon2014asymptotic}.
As in the reference, a numerical implementation of the propagators
(Eqs. \ref{eq:I-propagator}, \ref{eq:P-propagator}) requires three
elements: a suitable orbit integrator, a suitable interpolation procedure
of the function $f(\mathbf{z})$ from the computational mesh to the
orbit position, and a sufficiently accurate numerical quadrature of
integrals along orbits. For interpolation, we consider both second-order
multidimensional global \emph{and} local splines in a structured,
non-uniform logical mesh. The former is potentially more accurate
but more expensive, while the latter is significantly cheaper but
potentially less accurate. Here, we consider both to assess the practical
accuracy of the local spline interpolation approach for non-uniform
meshes (described in detail in \ref{sec:Local-spline-interpolation}).
Unless otherwise noted, results reported here have been obtained
with the local interpolation approach. We limit our interpolation
order to second order to ensure sufficient accuracy while minimizing
the chances of producing negative values, which is of concern since
the PDF tails can be quite rarefied and can be significantly distorted
by negative values. In the instances where negative interpolated values
for PDFs are found along orbits, we reset them to zero in the global
spline, and revert to first-order interpolation in the local spline.

Orbits and quadratures along orbits are computed simultaneously using
the high-order ODE integration package ODEPACK \citep{odepack}, which
features highly efficient, adaptive, high-order (up to 12th order)
numerical ODE integration routines. By treating the orbits and the
propagator integral on the same footing, ODEPACK is able to adapt
to structure in both the orbits and the integrands. The ODE set in
Eqs. \ref{eq:orbit_eq-E-mu} and \ref{eq:orbit_eq-gamma-mu} is not
stiff, and a simple Picard-implicit time-stepping integrator is sufficient.
The simulations presented in this study have been obtained with very
tight relative and absolute convergence tolerances for the ODE solver
($10^{-12}$), to minimize the impact of orbit integration errors
on the error convergence studies. The code is fully parallelized with
MPI, and orbits are threaded using OpenMP.

The numerical integration in Eqs. \ref{eq:I-propagator} and \ref{eq:P-propagator}
is performed along two semi-infinite orbit domains, $\eta'\in[0,\infty)$
and $\eta'\in(-\infty,0]$. Orbits are numerically truncated according
to error estimates based on the left-over integrals of the kernels,
which are estimated here (using similar techniques to those documented
in Ref. \citep{chacon2014asymptotic}) for a given tolerance $\varepsilon$
as:
\begin{eqnarray*}
\int_{\eta_{cut}}^{\infty}d\eta'\mathcal{G}(-\eta',D^{*})<\varepsilon & \Rightarrow & \eta_{cut}=2\sqrt{D^{*}}\xi_{cut}-1\,;\,\xi_{cut}\gtrsim\sqrt{|\ln(2\varepsilon\sqrt{\pi}\xi_{cut})|},\\
\int_{-\infty}^{-\eta_{cut}}d\eta'\mathcal{G}(-\eta',D^{*})<\varepsilon & \Rightarrow & \eta_{cut}=2\sqrt{D^{*}}\xi_{cut}+1\,;\,\xi_{cut}\gtrsim\sqrt{|\ln(2\varepsilon\sqrt{\pi}\xi_{cut})|},\\
\int_{\eta_{cut}}^{\infty}d\eta'\mathcal{K}(\eta',D^{*})<\varepsilon & \Rightarrow & \eta_{cut}=2\sqrt{D^{*}}\xi_{cut}\,;\,\xi_{cut}\gtrsim-\frac{1}{2\sqrt{D^{*}}}+\sqrt{\left|\ln\left(\varepsilon\sqrt{\pi}\xi_{cut}^{3/2}\left(\xi_{cut}+\frac{1}{2\sqrt{D^{*}}}\right)^{1/2}\right)\right|},\\
\int_{-\infty}^{-\eta_{cut}}d\eta'\mathcal{K}(\eta',D^{*})<\varepsilon & \Rightarrow & \eta_{cut}=2\sqrt{D^{*}}\xi_{cut}\,;\,\xi_{cut}\gtrsim\frac{1}{2\sqrt{D^{*}}}+\sqrt{\left|\ln\left(\varepsilon\sqrt{\pi}\xi_{cut}^{3/2}\left(\xi_{cut}-\frac{1}{2\sqrt{D^{*}}}\right)^{1/2}\right)\right|}.
\end{eqnarray*}
In these expressions, $\xi_{cut}$ can be found very quickly for a
prescribed tolerance $\varepsilon$ once at the beginning of the simulation
with a few Picard iterations. The asymmetry in the integral bounds
with respect to the sign of $\eta$ originates in the advective character
of the $\mathcal{G}$ and $\mathcal{K}$ kernels. It is worth noting
that the integral bound cutoff in the original integration variable
$\tau=\Delta t\eta/\epsilon$ is roughly proportional to $\Delta t\sqrt{D^{*}}/\epsilon$,
indicating that the cost of the orbit integrals will grow accordingly.

We finalize this section commenting on the conservation properties
of the semi-Lagrangian approach. The nonlinear collisional source
is fully conservative in mass, momentum, and energy \citep{daniel2020-irfp}.
The Lagrangian step is also conservative in the continuum, but in
principle not in the discrete. We enforce mass conservation errors
(typically of 1 part in $10^{7}$ after the orbit integration) by
renormalizing our PDF after the Lagrangian update to the integrated
mass before the update. This strategy corrects for long-term error
accumulation. We do not correct for errors in momentum or energy conservation.
We will demonstrate numerically in the next section that these remain
small, $\sim\mathcal{O}(10^{-9})$, in typical simulations.

\section{\label{sec:Numerical-results}Numerical results}

We consider several tests in increasing order of complexity to demonstrate
the asymptotic properties and correctness of the semi-Lagrangian algorithm.
We begin with a 0D-2P Dreicer generation test problem, which allows
us to compare the semi-Lagrangian implementation to a purely Eulerian
one. We demonstrate that both agree very well with each other, and
with documented results elsewhere in the literature. Next, following
Ref. \citep{mcdevitt2019runaway}, we consider the runaway-electron
Dreicer generation rate in a 2D axisymmetric circular tokamak geometry.
We use this nontrivial geometry to demonstrate the conservation properties
of the semi-Lagrangian scheme, and verify our implementation against
the results in the reference, demonstrating excellent agreement. We
also use this test to demonstrate the advertised asymptotic properties
of the scheme, namely, uniform convergence in $\Delta t/\epsilon$,
and the asymptotic convergence to the limit (bounce-averaged \citep{nilsson2015kinetic})
solution.

\subsection{0D-2P Dreicer generation problem}

\begin{figure}
\begin{centering}
\includegraphics[width=3in]{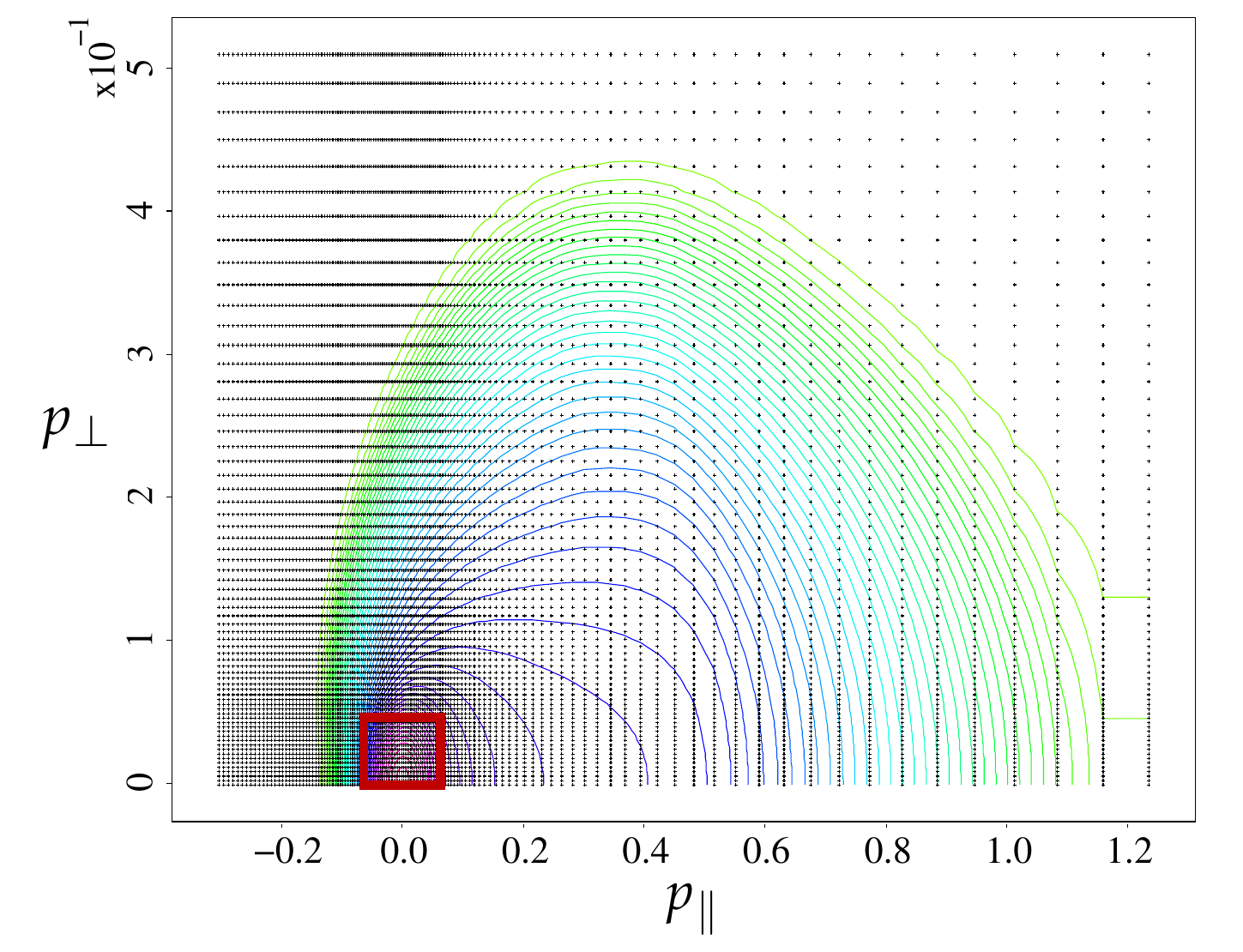}
\par\end{centering}
\caption{\label{fig:Sample-hybrid-uniform-geometric}Sample hybrid uniform-geometric
mesh employed in the 0D-2P Dreicer generation problem, along with
Eulerian-$E$ solution at $t=900\tau_{c}^{th}$. This particular mesh
has 128$\times$64 points, with a uniform mesh (red box) of 32$\times$16
points resolving the bulk Maxwellian in $p_{\parallel}\in(-3.5v_{th}/c,3.5v_{th}/c)$
and $p_{\perp}\in(0,3.5v_{th}/c)$ (with $3.5v_{th}/c\approx0.05$),
and expanding geometrically from there to the domain limits.}
\end{figure}

We consider first a 0D-2P Dreicer runaway generation problem with
no magnetic field. The purpose of this test is to verify the simplest
semi-Lagrangian scheme (where only the electric field is treated in
a Lagrangian manner) against the fully Eulerian implementation in
Ref. \citep{daniel2020-irfp}. The orbit equations for the semi-Lagrangian
algorithm in this case simply become:
\[
\frac{dp_{\parallel}}{d\tau}=-E_{\parallel}.
\]
Unless otherwise specified, we consider a Dreicer-normalized electric
field $E_{\parallel}=0.06$, and a plasma temperature of 100 eV ($\Theta_{0}=(v_{th}/c)^{2}=1.95\times10^{-4}$).
The normalized momentum domain is $p_{\parallel}\in(-0.3,1.2)$, $p_{\perp}\in(0,0.5)$,
with a 512$\times$128 hybrid uniform-geometric mesh, with a 256$\times$32
uniform mesh resolving the thermal bulk in $p_{\parallel}\in(-5v_{th}/c,5v_{th}/c)$
and $p_{\perp}\in(0,5v_{th}/c)$ (with $5v_{th}/c\approx0.07$), and
expanding geometrically from there to the outer domain limits with
the remaining mesh points. The domain limits in $p_{\parallel}$ were
chosen such that a sufficient runaway tail can be measured by our
Dreicer production diagnostic. The timestep used is $\Delta t=5\times10^{-7}$
in relativistic units ($\approx0.18$ thermal collision times) for
the Eulerian simulation (which we term ``Eulerian-$E$'' to highlight
the fact that the electric field is treated on the mesh), and $\Delta t=1\times10^{-7}$
and $D^{*}=0.05$ for the semi-Lagrangian ones (which we term ``Lagrangian-$E$''
to highlight the fact that the electric field is integrated along
orbits). As we show below, larger timesteps in the Lagrangian-$E$
treatment result in significant deviation of the Dreicer generation
rate vs. the Eulerian result owing to the splitting introduced between
collisions and acceleration terms. We run the simulation to $t=2.5\times10^{-3}$,
or 900 thermal collision times. A sample hybrid uniform-geometric
128x64 mesh on a similar domain is provided in Fig. \ref{fig:Sample-hybrid-uniform-geometric},
along with the Eulerian solution at 900 $\tau_{c}^{th}$.
 The Dreicer production rate (in relativistic collision time units)
$\sigma_{rel}$ is measured as:
\[
\sigma_{rel}=\frac{\partial_{t}n_{RE}}{n_{0}-n_{RE}},
\]
where $n_{0}$ is the initial electron density, and $n_{RE}$ is the
runaway electron population, which is measured from the solution for
electron velocities exceeding $p_{cutoff}\approx$ 25 $v_{th}/c\approx0.35$
in the positive $p_{\parallel}$ domain, i.e.:
\[
n_{RE}=\int_{p_{\parallel,cutoff}}^{\infty}dp_{\parallel}\int_{p_{\perp,cutoff}}^{\infty}dp_{\perp}p_{\perp}f,
\]
with $p_{\parallel,cutoff}^{2}+p_{\perp,cutoff}^{2}\geq p_{cutoff}^{2}$
and $p_{\parallel,cutoff}>0$.

\begin{figure}
\begin{centering}
\includegraphics[width=6in]{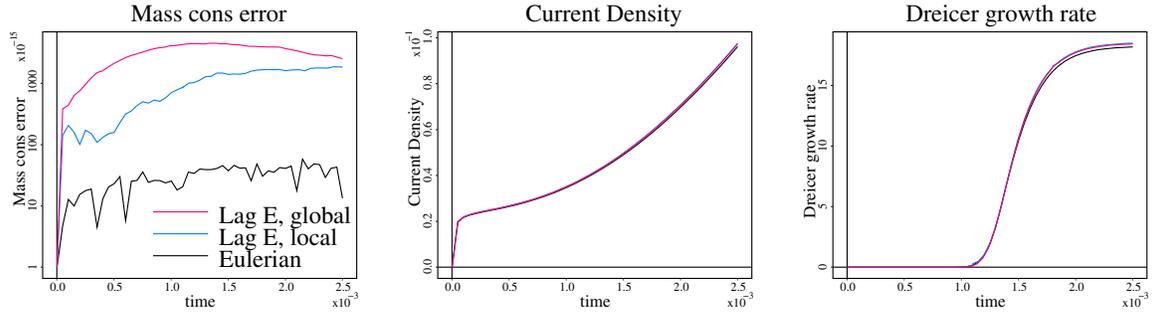}
\par\end{centering}
\caption{\label{fig:semilag-vs-eulerian}Comparison of the Eulerian and semi-Lagrangian
time histories of mass conservation error, Dreicer current density
generation, and Dreicer growth rate for the 0D-2P Dreicer generation
problem, demonstrating excellent agreement.}
\end{figure}
Figure \ref{fig:semilag-vs-eulerian} depicts the comparison of time
traces of mass conservation error, kinetic energy, current density
carried by runaway electrons, and Dreicer growth rate from both Eulerian
and semi-Lagrangian implementations (the latter using both global
and local interpolation of the propagator integrands), demonstrating
excellent agreement in all cases. For these cold plasma conditions,
we can compare the Dreicer generation rate against the results in
Ref. \citep{kulsrud1973runaway} (which were obtained with a non-relativistic
collisional treatment) to find excellent agreement ($\sigma_{rel}=18.3$
in the figure vs. $\sigma_{rel}=\sigma(c/v_{th})^{3}=18.5$, with
$\sigma=5\times10^{-5}$, in the reference). This suggests that the
results in Fig. \ref{fig:semilag-vs-eulerian} are converged. The
semi-Lagrangian 0D run with local interpolation is a factor of two
faster (in wall-clock time) than the global-spline run. In these results,
the Lagrangian-$E$ treatment uses a timestep $5\times$ smaller than
the Eulerian-$E$ run for accuracy. 
\begin{figure}
\begin{centering}
\includegraphics[width=2in]{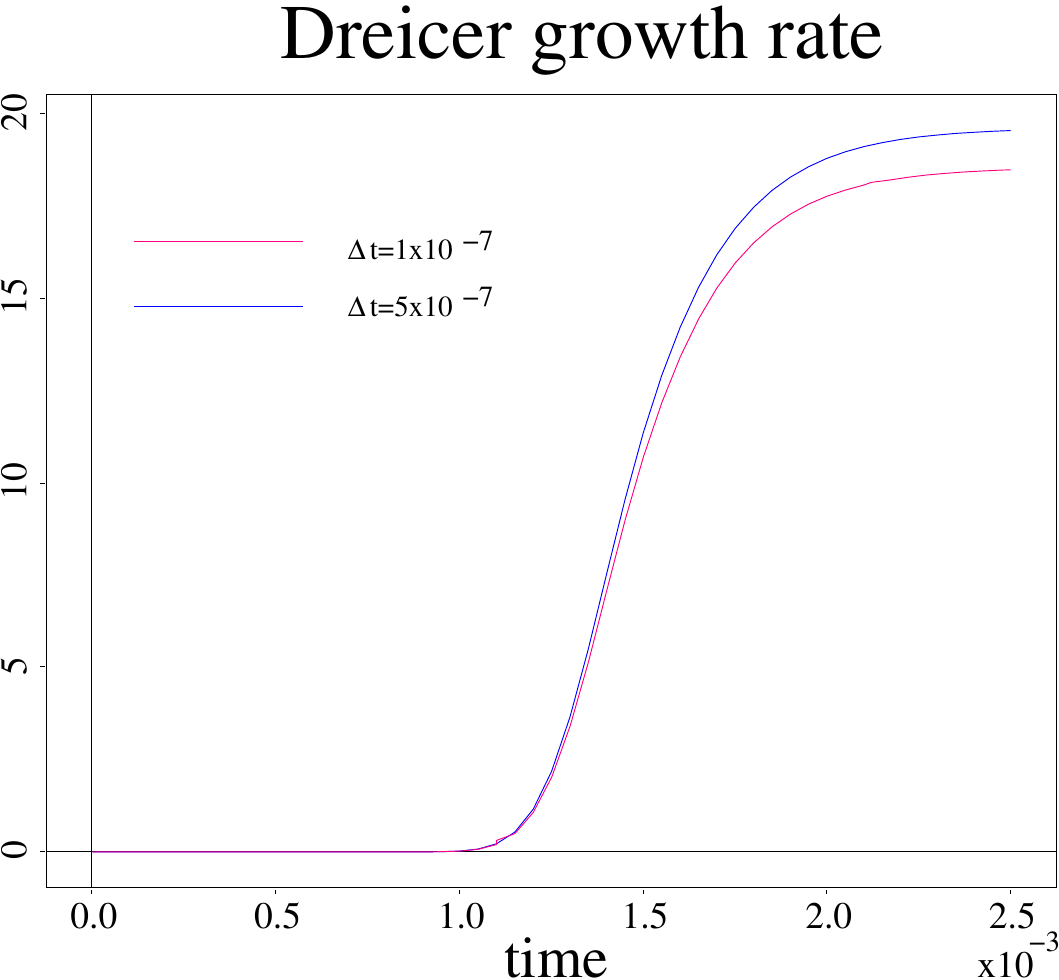}
\par\end{centering}
\caption{\label{fig:lag-E-dt-comp}Impact of timestep size in the accuracy
of the computation of the 0D-2P Dreicer growth rate with the Lagrangian-$E$
treatment. The larger timestep ($\Delta t=5\times10^{-7}$, which
gives an accurate solution with the Eulerian-$E$ treatment) results
in significant overestimation of the Dreicer growth rate by the Lagrangian-$E$
treatment. This outcome is insensitive to the choice of interpolation
in the semi-Lagrangian algorithm.}
\end{figure}
Figure \ref{fig:lag-E-dt-comp} illustrates the accuracy impact in
the Dreicer growth rate when a larger time step is used in the Lagrangian-$E$
simulation. For the same timestep as in the Eulerian-$E$ treatment
($\Delta t=5\times10^{-7}$), the semi-Lagrangian simulation introduces
significant error in the Dreicer growth rate, indicating that the
Eulerian-$E$ treatment is preferable from an accuracy standpoint
for a given timestep. Results in the next section confirm that this
conclusion applies to 2D-2P simulations as well.

\subsection{2D-2P Dreicer generation in axisymmetric tokamak}

In principle, our proposed algorithm can deal with general three-dimensional
magnetic field configurations. In this study, we consider 2D axisymmetric
toroidal topologies, of recent interest \citep{mcdevitt2019runaway}.
In such topologies, the magnetic field can be described by an axisymmetric
poloidal flux function $\Psi$ (i.e., independent of the toroidal
angle $\phi$), and a toroidal magnetic field $B_{\phi}$ as:
\[
\mathbf{B}=\nabla\phi\times\nabla\Psi+RB_{\phi}\nabla\phi.
\]
Clearly, $\mathbf{B}$ is tangential to flux surfaces, and the orbit
dynamics can be described by a poloidal angle $\theta$ along a given
flux surface, $\Psi$. Therefore, the spatial dependence of the PDF
can be described uniquely by the coordinates ($\Psi,\theta$), and
we have:
\[
f=f(\Psi,\theta,p_{\parallel},p_{\perp}).
\]

\begin{figure}
\begin{centering}
\includegraphics[width=3in]{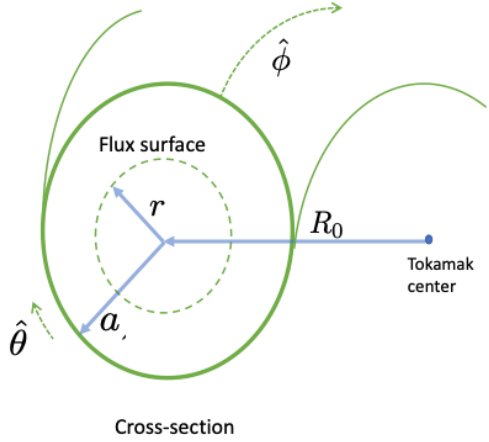}
\par\end{centering}
\caption{\label{fig:Sketch-of-axisymmetric}Sketch of axisymmetric cylindrical
tokamak geometry.}
\end{figure}
We consider the same geometry as in Ref. \citep{mcdevitt2019runaway},
i.e., a circular tokamak geometry of normalized radius $a=1$ (see
Fig. \ref{fig:Sketch-of-axisymmetric}), with the q-profile given
by: 
\[
q(r)=\frac{rB_{\phi}}{RB_{\theta}}=2.1+2r^{2}.
\]
Here, $r$ is the toroidal minor radius, which in this configuration
is a flux function and therefore can replace $\Psi$ as a coordinate.
The corresponding poloidal and toroidal magnetic field components
($B_{\theta},B_{\phi})$ are given by:
\begin{eqnarray*}
B_{\phi} & = & \frac{R_{0}B_{0}}{R}=\frac{B_{0}}{1+\frac{r}{R_{0}}\cos(\theta)},\\
B_{\theta} & = & \frac{rB_{\phi}}{R_{0}q(r)},
\end{eqnarray*}
with $R_{0}=3a$ the toroidal major radius \citep{mcdevitt2019runaway}
and $B_{0}$ a reference magnetic field (which can be set to unity
without loss of generality, as it does not enter the final orbit equations).
The corresponding magnetic field magnitude is:
\[
B=\sqrt{B_{\theta}^{2}+B_{\phi}^{2}}=B_{\phi}\sqrt{1+\left(\frac{r}{qR_{0}}\right)^{2}}.
\]
In circular geometry, the orbit equations in relativistic units (Eq.
\ref{eq:orbit_eq-E-mu} or Eq. \ref{eq:orbit_eq-gamma-mu}, depending
on the representation) can be simplified as:
\begin{eqnarray}
\frac{d\theta}{d\tau} & = & \frac{d\mathbf{x}}{d\tau}\cdot\frac{\partial\theta}{\partial\mathbf{x}}=\frac{d\mathbf{x}}{d\tau}\cdot\frac{\boldsymbol{\theta}}{r}=\frac{p_{\parallel}}{\gamma}\frac{B_{\theta}}{rB},\label{eq:theta_update}\\
\frac{dp_{\parallel}}{d\tau} & = & \left[-E_{\parallel}\right]-\frac{p_{\perp}^{2}}{2\gamma}\frac{\mathbf{b}\cdot\nabla B}{B}=\left[-E_{\parallel}\right]-\frac{p_{\perp}^{2}}{2\gamma}\frac{B_{\theta}\partial_{\theta}B}{rB^{2}},\label{eq:ppar_update}\\
\frac{dp_{\perp}}{d\tau} & = & \frac{p_{\perp}p_{\parallel}}{2\gamma}\frac{\mathbf{b}\cdot\nabla B}{B}=\frac{p_{\perp}p_{\parallel}}{2\gamma}\frac{B_{\theta}\partial_{\theta}B}{rB^{2}},\label{eq:pperp_update}
\end{eqnarray}
where:
\[
\frac{B_{\theta}}{B}=\frac{r}{\sqrt{(R_{0}q)^{2}+r^{2}}}\,\,;\,\,\frac{\partial_{\theta}B}{rB}=\frac{\sin(\theta)}{R_{0}+r\cos(\theta)}.
\]
The parallel (toroidal) electric field $E_{\parallel}$ is defined
from a gradient, and therefore in toroidal geometry satisfies:
\[
E_{\parallel}=\frac{E_{0}R_{0}}{R}=\frac{E_{0}}{1+\frac{r}{R_{0}}\cos(\theta)},
\]
with $E_{0}$ a reference electric field in relativistic (Connor-Hastie)
units.

\subsubsection{Conservation tests of the semi-Lagrangian algorithm with nonlinear
collisions in the absence of external drivers}

\begin{figure}
\begin{centering}
\includegraphics[width=2.2in,height=2in]{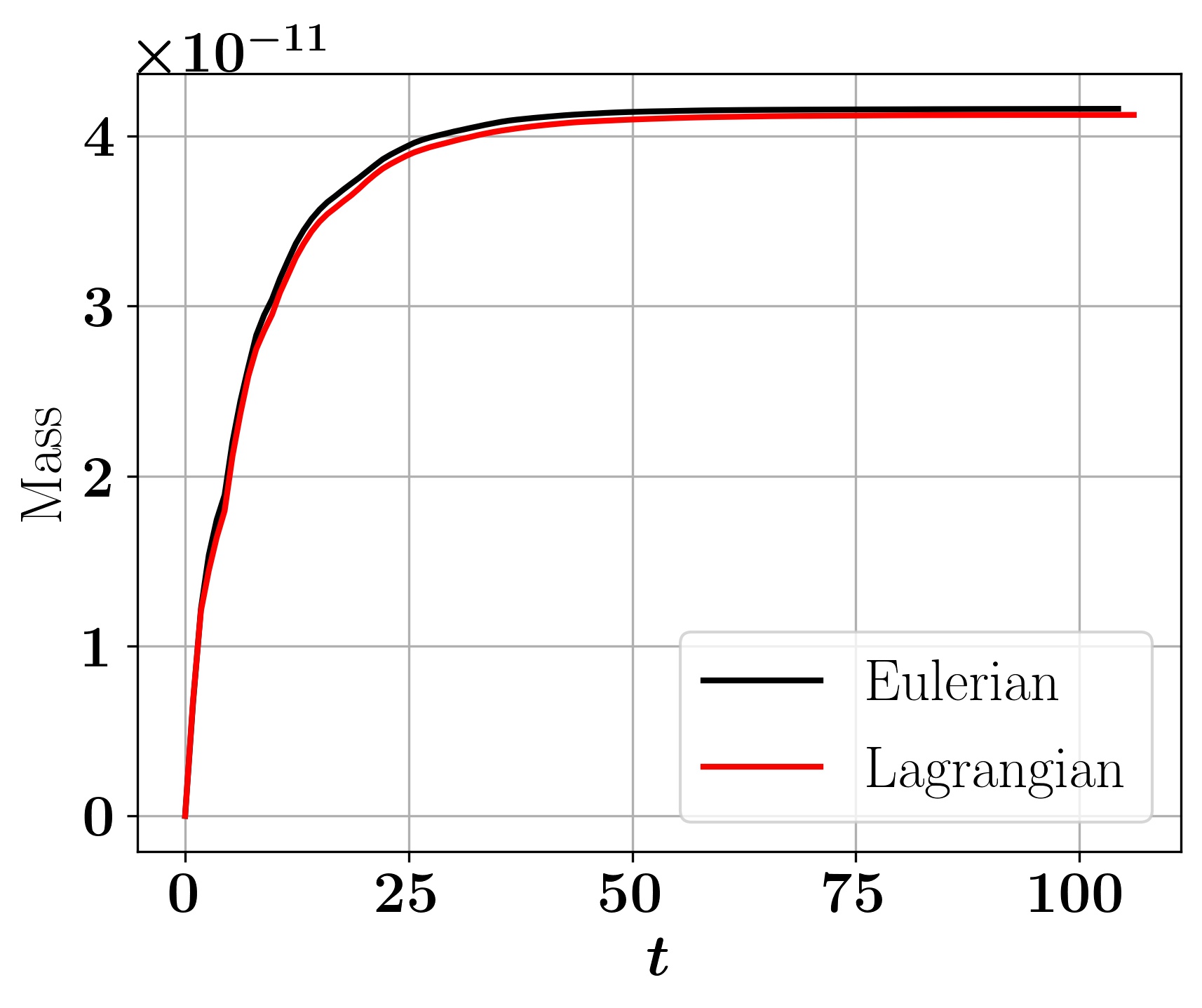}\includegraphics[width=2.2in,height=1.9in]{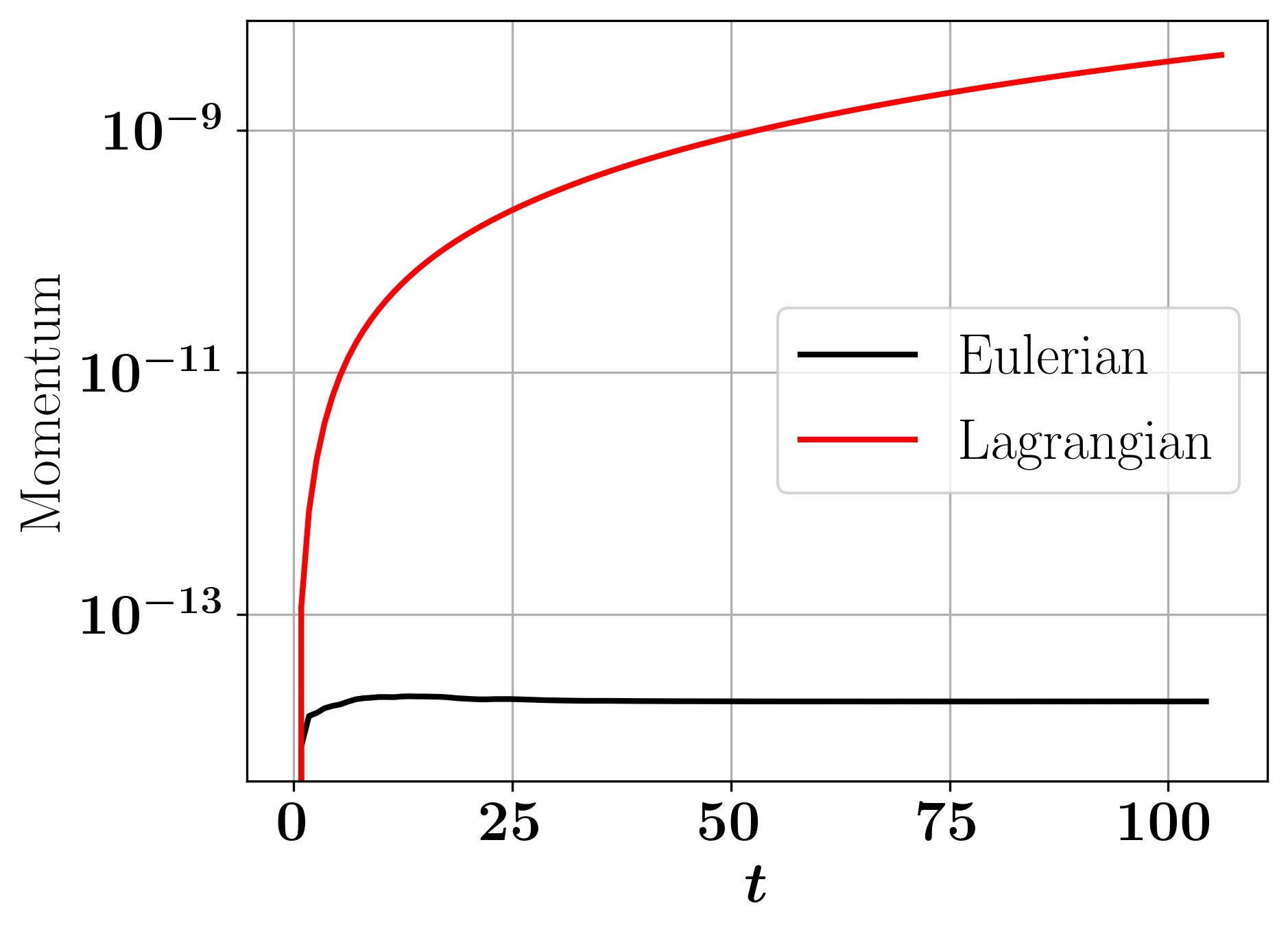}\includegraphics[width=2.2in,height=1.9in]{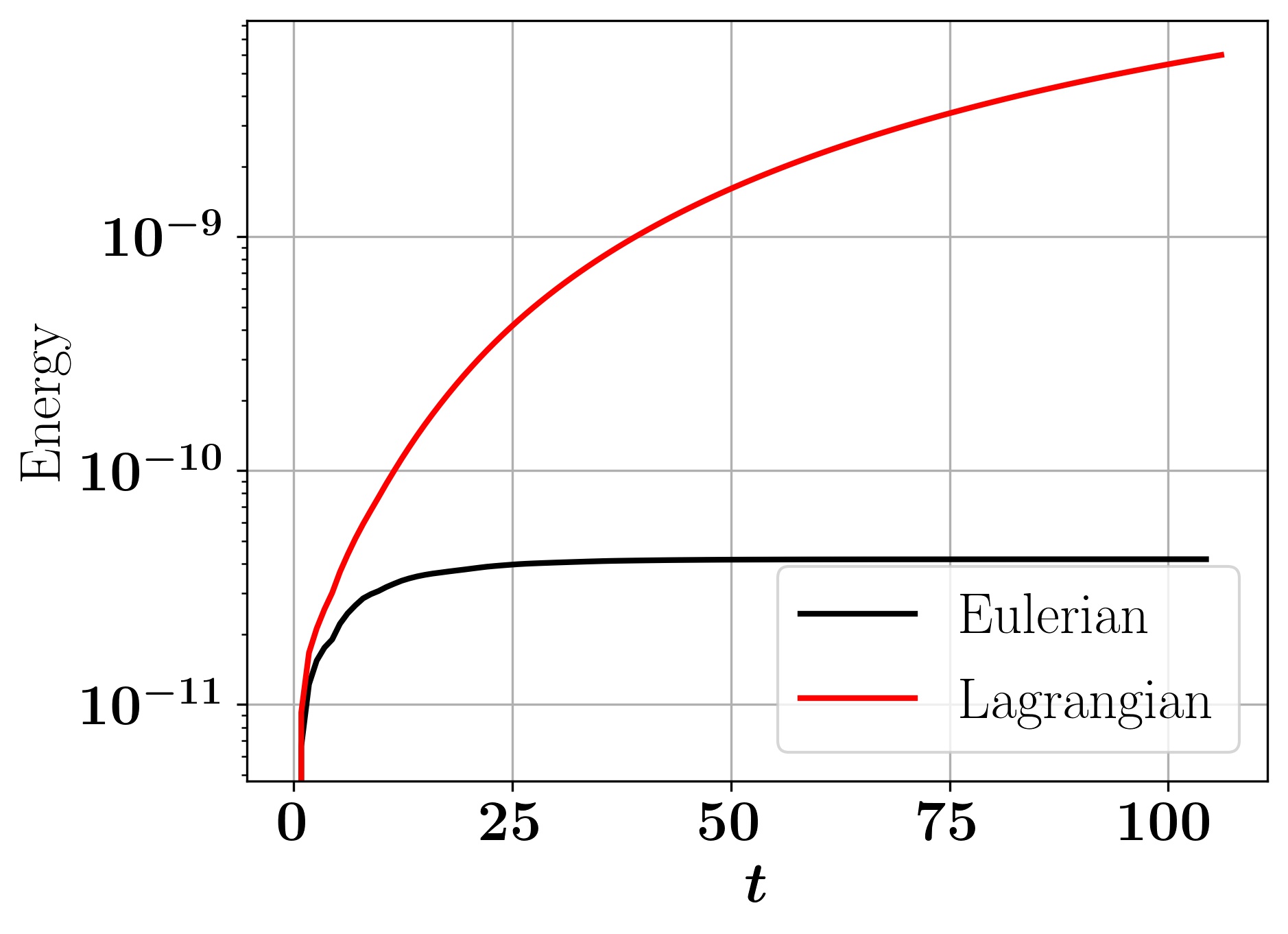}
\par\end{centering}
\caption{\label{fig:conservation}Off-axis ($r/a=0.4$) thermal equilibrium
test with no external driver, demonstrating excellent conservation
properties of the semi-Lagrangian algorithm (red line) for mass, momentum,
and energy. Conservation properties for the Eulerian discretization
on-axis ($r=0$, for which the angular coordinate is ignorable, and
therefore features no spatial dependence) are provided as a reference
(black line).}
\end{figure}
We begin by testing the conservation properties of the semi-Lagrangian
algorithm without external drivers such as an electric field, but
including fully nonlinear and conservative electron-electron collisions.
We recall that the collisional treatment is fully conservative when
solved nonlinearly, as has been documented elsewhere \citep{daniel2020-irfp}.
The test therefore highlights the conservation errors introduced by
the transport (collisionless) component. We consider an initial condition
in thermal equilibrium for a temperature of 100 eV and $E_{\parallel}=0$
(i.e., no external driver is present). Simulations are performed on
a $8\times128\times64$ grid with domain size in $p_{\parallel}\in(-0.3,0.3)$
and $p_{\perp}\in(0,0.3)$. Eulerian simulations are performed on-axis
($r=0$), for which the angular direction is ignorable, and are provided
here as a reference. Lagrangian simulations were performed off-axis
($r/a=0.4$), using the local interpolation scheme for 100 thermal
collisional steps. Results are reported in Fig. \ref{fig:conservation},
and demonstrate the ability of the scheme to conserve invariants in
less than one part in $10^{8}$, thus maintaining thermal equilibrium
to high precision.

\subsubsection{Verification tests}

We consider the plasma radii
\[
\frac{{r}}{a}=0,0.2,0.4,0.6,0.8,
\]
and plasma temperatures of 100eV, 1 keV, and 10 keV. For consistency
with Ref. \citep{mcdevitt2019runaway}, we consider the linearized
form of the relativistic collision operator, in which the background
electron population does not evolve. To reproduce the results in the
reference, the Coulomb logarithm in the collision operator is computed
for the given density and temperature values according to well-known
formulas \citep{huba2009nrl}. We measure the Dreicer electron production
growth rate for an effective ion charge $Z_{\mathrm{{eff}}}=1$ and
an electric field $E_{0}=0.06\,E_{D}/E_{CH}=0.06(c/v_{th})^{2}$.
Considering a temperature-dependent Coulomb logarithm, $R_{0}=6$
m in physical units, and a density of $n=2\times10^{14}$ cm$^{-3}$,
we obtain $\epsilon_{rel}=1.45\times10^{-6},\,1.73\times10^{-6},\,2\times10^{-6}$
(corresponding to $\epsilon=38,\,0.45,\,0.0052$) for 100 eV, 1 keV
and 10 keV, respectively. \textcolor{black}{As before, the simulations
span a total time of 900-1000 $\tau_{c}^{th}$.}

We consider both a Lagrangian-$E$ treatment (Eq. \ref{eq:rDKE-nc-E-mu})
and an Eulerian-$E$ treatment (Eq. \ref{eq:rDKE-nc-gamma-mu}).\textcolor{black}{{}
From the results of the stability analysis, we choose $D^{*}=0.2$
for all Eulerian-$E$ cases and $D^{*}=0.05$ for all Lagrangian-$E$
cases. }Timesteps with the Eulerian-$E$ treatment for the colder
plasmas (100 eV and 1 keV) are $\Delta t\approx0.2$ $\tau_{c}^{th}$
{[}or, in relativistic (code) units, $\Delta t=0.2(v_{th}/c)^{3}${]},
while for the hot case (10 keV) we use $\Delta t\approx0.07$ $\tau_{c}^{th}$.
As observed before, the Lagrangian-$E$ treatment requires timesteps
a factor of 2 or so smaller than those with an Eulerian-$E$ treatment
for good agreement, owing to the additional splitting error introduced
when dealing with the electric-field acceleration in the orbit integrals.
The nonlinear tolerance is set to $10^{-8}$ and background potentials
are solved at the beginning of the simulation with a linear relative
tolerance of $10^{-10}.$

We use a geometrically stretched momentum grid of $512\times128$
for the cold case (100 eV), while we use $256\times128$ for the hotter
cases. Grid spacing at the origin is $\Delta p=5\times10^{-4},\,3\times10^{-3},\,5\times10^{-2}$
for 100 eV, 1 keV, 10 keV, respectively. We have also confirmed that
the results are reproducible with a $256\times128$ hybrid grid with
the uniform domain of $128\times64$ centered around $5$ $v_{th}/c.$
The normalized momentum domain is adjusted for the different temperatures:
$p_{\parallel}\in(-0.3,1.2)$, $p_{\perp}\in(0,0.5)$ for 100 eV,
$p_{\parallel}\in(-1.0,2.5)$, $p_{\perp}\in(0,1)$ for 1 keV, and
$p_{\parallel}\in(-3,7)$, $p_{\perp}\in(0,5)$ for 10 keV. As before,
the domain limits in $p_{\parallel}$ were chosen such that a sufficient
runaway tail is formed for Dreicer production measurements, and runaway
electron populations are measured for electron velocities exceeding
$\approx$ 12-25 $v_{th}/c$. The poloidal dimension employs a uniform
mesh with 64, 64, and 16 points for 100 eV, 1 keV, and 10 keV, respectively.
We initialized the electron distribution in the momentum space with
the relativistic Maxwell-Juttner distribution except for the cold
case of 100 eV, where we used a non-relativistic Maxwell distribution.

\begin{figure}
\begin{centering}
\subfloat[Eulerian electric-field treatment]{\begin{centering}
\includegraphics[width=3in,height=2.25in]{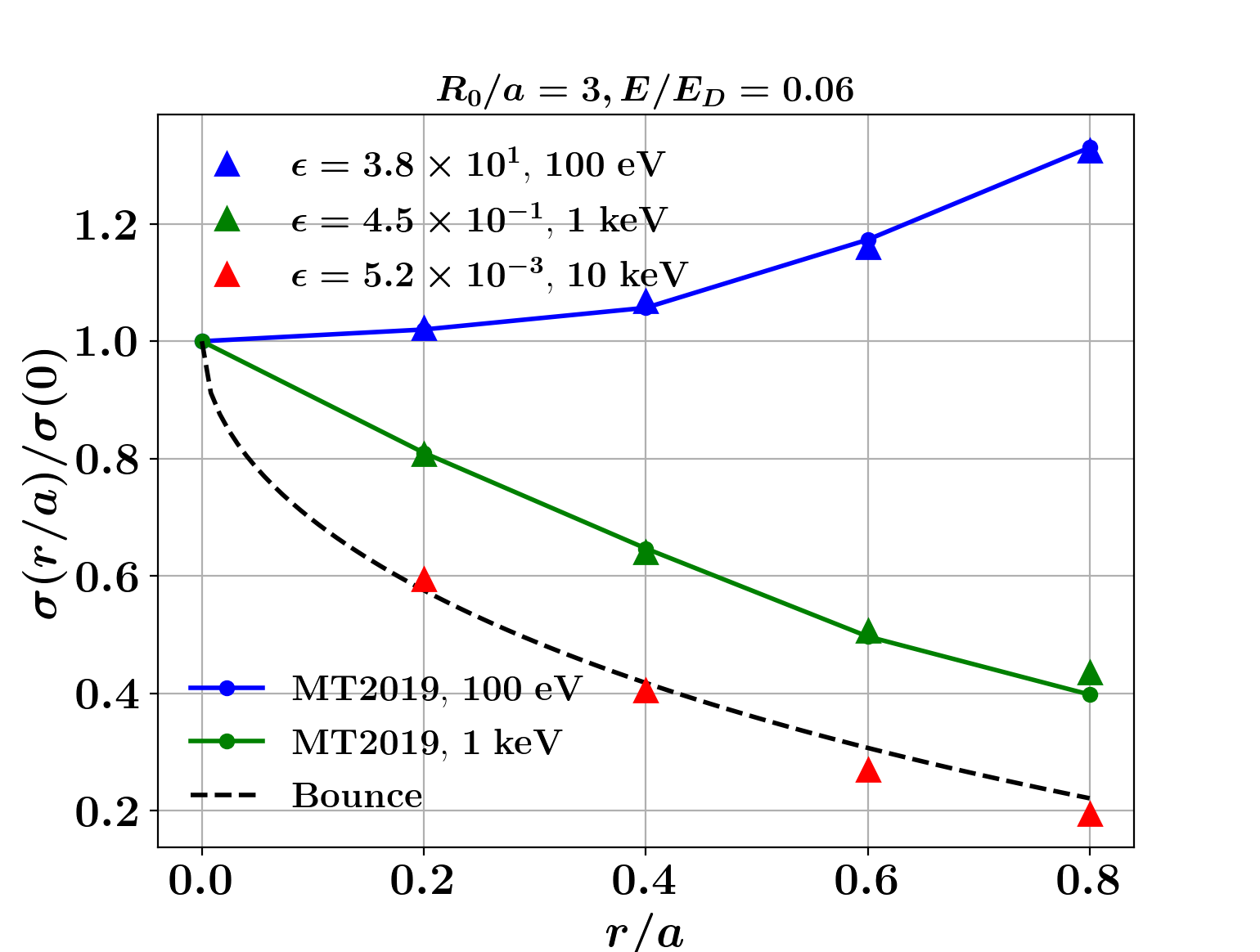}
\par\end{centering}
}\subfloat[Lagrangian electric-field treatment]{\begin{centering}
\includegraphics[width=3.2in,height=2.25in]{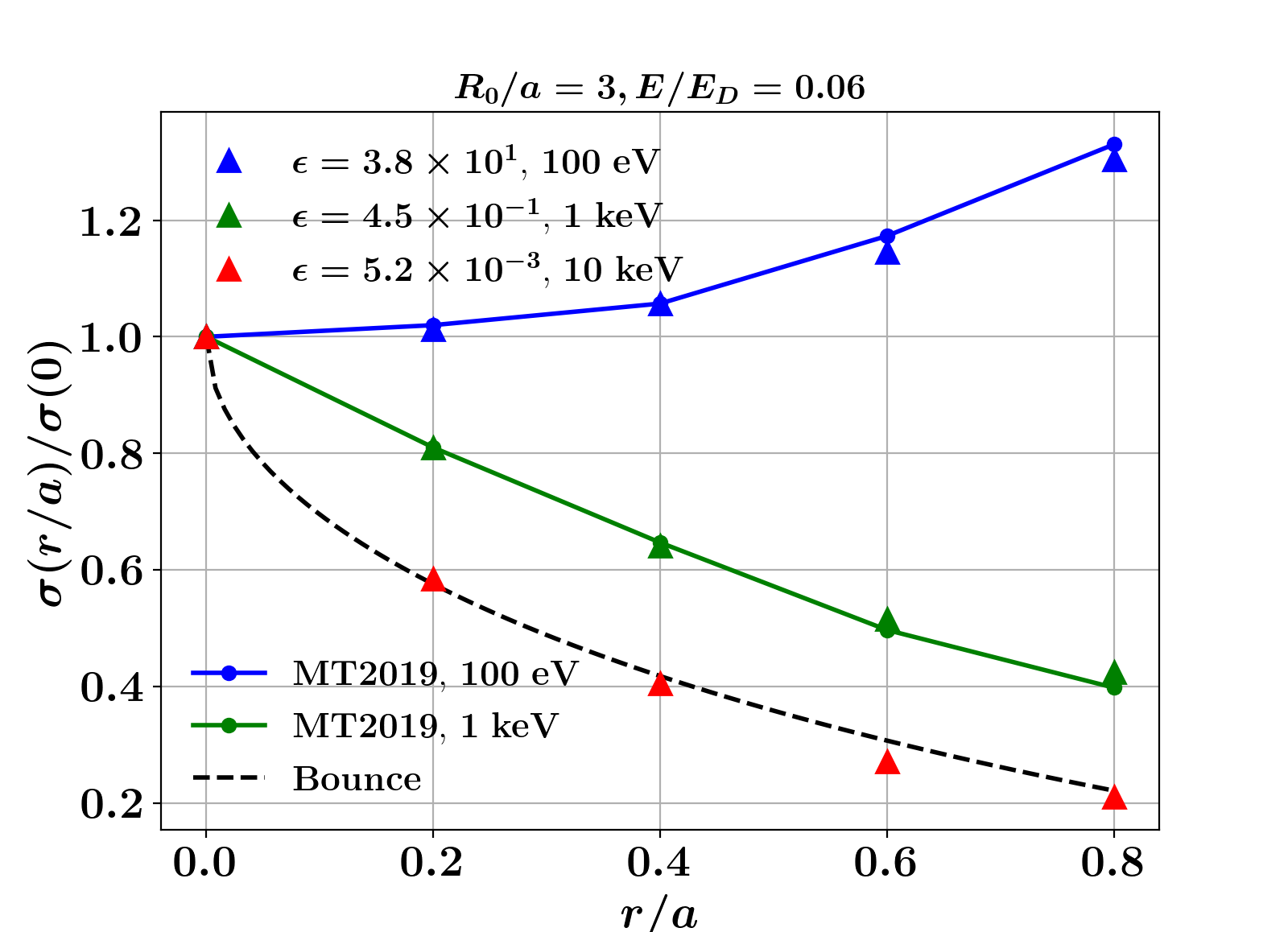}
\par\end{centering}
}
\par\end{centering}
\caption{\label{fig:Dreicer-generation-rates}Comparison of the relative change
of Dreicer generation rates vs. the $r=0$ value along radius in the
axisymmetric cylindrical tokamak geometry with results in Fig. 4 in
Ref. \citep{mcdevitt2019runaway}, with the electric field treated
(a) either on the mesh with $D^{*}=0.2$ or (b) in the orbit integral
with $D^{*}=0.05$. The bounce-averaged solution is from Ref. \citep{nilsson2015kinetic}.
Note that the asymptotic parameter $\epsilon$ in the left figure
ranges from 38 to 0.005 depending on the plasma temperature, illustrating
the need for a uniformly convergent algorithm. These plots demonstrate
that the results are not only largely independent of the electric-field
treatment, but also on the value of $D^{*}$ (provided it is chosen
in the stable range).}
\end{figure}
The key result of this study is depicted in Fig. \ref{fig:Dreicer-generation-rates},
which compares the relative Dreicer generation rates for various plasma
temperatures as a function of plasma radius to the results in Ref.
\citep{mcdevitt2019runaway}. Both Lagrangian-$E$ and Eulerian-$E$
approaches demonstrate excellent agreement with published data for
all plasma temperatures, illustrating the ability of the method to
treat all relevant asymptotic regimes. In particular, it demonstrates
that our implementation is not just asymptotic-preserving (i.e., able
to capture the $\epsilon\rightarrow0$ limit), but in fact uniformly
convergent in $\Delta t/\epsilon$.

\begin{figure}
\begin{centering}
\includegraphics[height=2.25in]{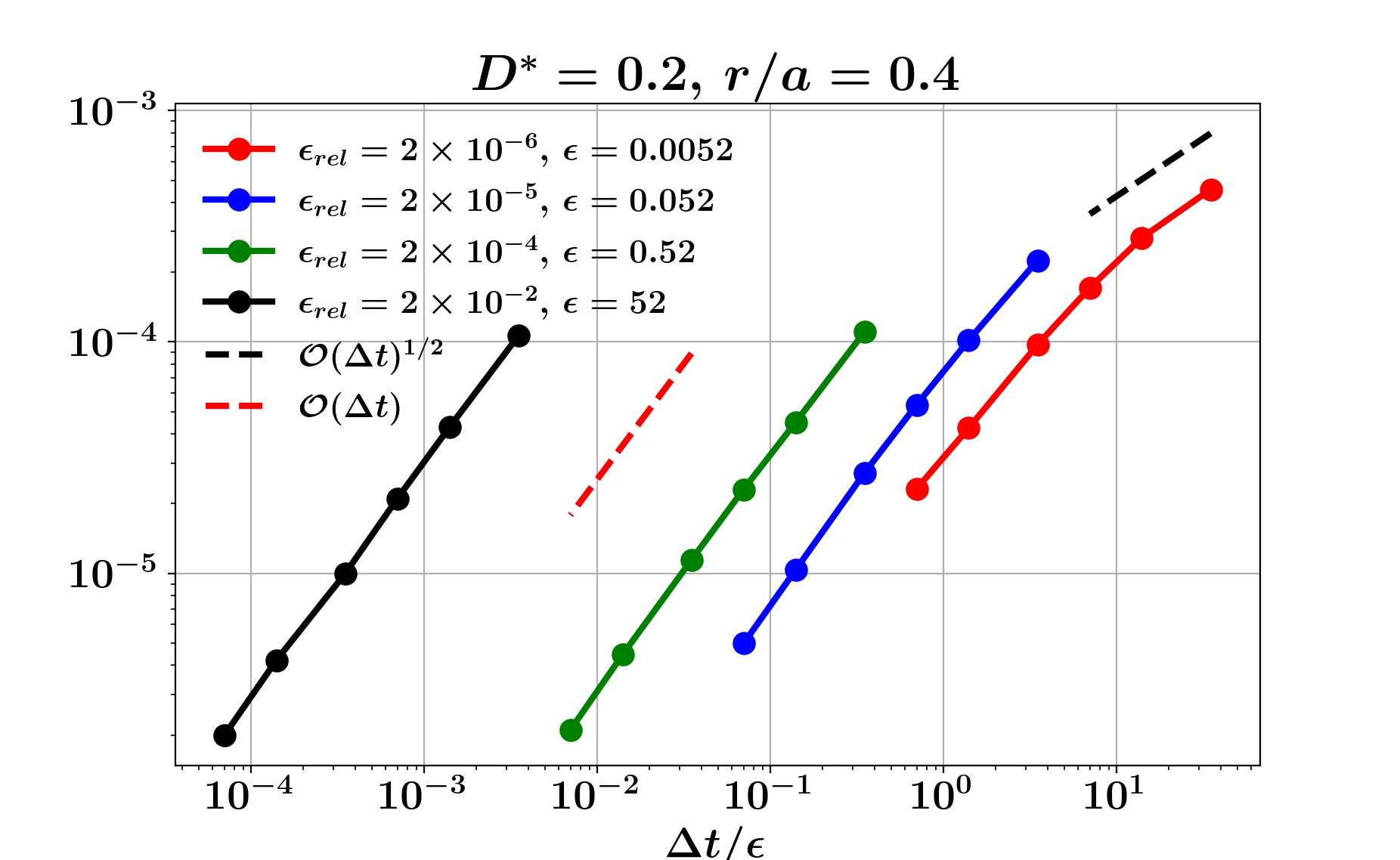}~~\includegraphics[height=2.15in]{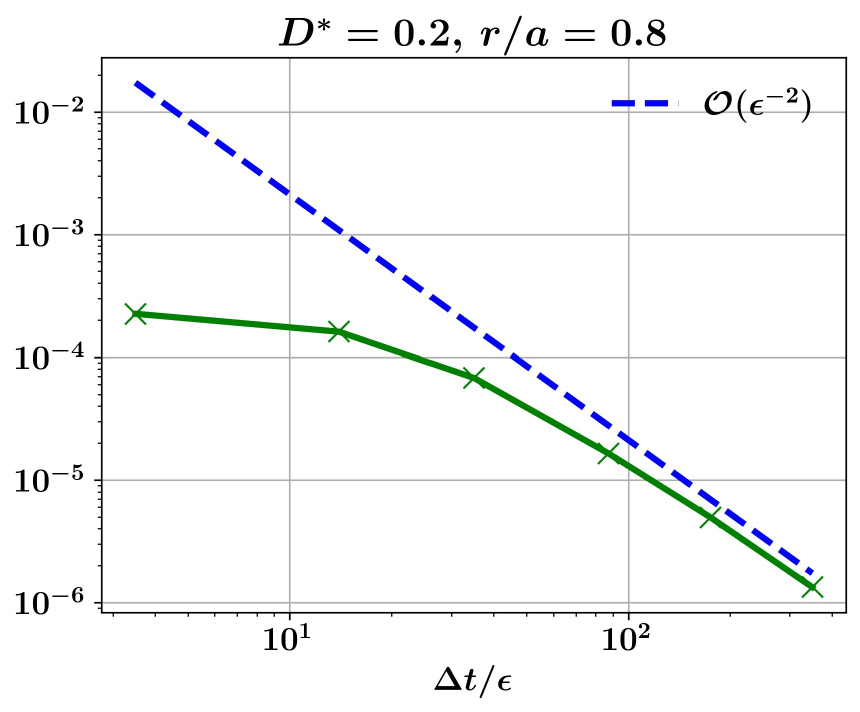}
\par\end{centering}
\caption{\label{fig:convergence}Convergence study results for the cylindrical
tokamak 10 keV plasma in terms of $\Delta t/\epsilon$ showing first-order
temporal convergence for $\Delta t/\epsilon$ small (left) and $\epsilon^{2}$
for $\Delta t/\epsilon$ large (right), with particular choices of
$D^{*}$ and $r/a$ as indicated in the title figures. Results are
obtained with the Eulerian electric-field treatment.\textcolor{red}{}}
\end{figure}
This point is further demonstrated in Fig. \ref{fig:convergence},
where we show error convergence studies in terms of both $\Delta t$
and $\epsilon$ for the 10 keV case with an Eulerian-$E$ treatment.
The error is computed as a volume-averaged root-mean-square of the
solution difference,
\[
\|f-f_{ref}\|_{2}=\left(\frac{{1}}{\pi L_{p_{\parallel}}L_{p_{\perp}}^{2}}\sum_{i=1}^{N_{\theta}}\sum_{j=1}^{N_{\parallel}}\sum_{k=1}^{N_{\perp}}(f-f_{ref})^{2}p_{\perp}\Delta p_{\parallel}\Delta p_{\perp}\Delta\theta\right)^{1/2},
\]
where $f_{ref}$ is the reference solution. The figure on the left
was obtained for various $\epsilon$ values by varying the time step
$\Delta t$. Errors for a given $\epsilon$ are computed at a final
time that was 10 times the largest time step, and are measured with
respect to a reference solution employing the same mesh resolution
and a timestep that is an order of magnitude smaller than the smallest
timestep considered. Errors in the figure on the right employ a single
time step of $\Delta t=0.2\tau_{c}^{th}$, and $\epsilon=(5.2,10.45,20.9,52.2,130.6,522.7)\times10^{-4}$.
Errors are computed with respect to a reference solution obtained
on the same mesh and timestep and $\epsilon=5.2\times10^{-5}$. In
both these figures, \textcolor{black}{$\epsilon$ is varied by artificially
changing} the plasma radius $R_{0}$ \textcolor{black}{for constant
plasma temperatur}e. Figure \ref{fig:convergence}-left demonstrates
that the algorithm is first-order accurate in $\Delta t$ \textcolor{black}{as
long as $\Delta t/\epsilon<\mathcal{O}(1)$, while Fig. }\ref{fig:convergence}\textcolor{black}{-right
shows that the convergence of the algorithm in the large $\Delta t/\epsilon$
asymptotic regime scales as $\epsilon^{2}$. These results confirm
the analysis in Sec. \ref{subsec:Asymptotic-conv}.}

\begin{table}
\centering{}\caption{\label{tab:D*-insensitivity}Relative Dreicer growth rate $\sigma/\sigma(0)$
at $r/a=0.6$ for various temperatures and $D^{*}$ values using the
Eulerian-$E$ treatment, demonstrating insensitivity of this physical
figure of merit to $D^{*}$ .}
\begin{tabular}{|c|c|c|}
\hline 
$D^{*}$ & 0.2 & 1.0\tabularnewline
\hline 
\hline 
100 eV & 1.157 & 1.160\tabularnewline
\hline 
1 keV & 0.501 & 0.507\tabularnewline
\hline 
\end{tabular}
\end{table}

A desirable property is the insensitivity of the physics to the numerical
parameter $D^{*}$. Figure \ref{fig:Dreicer-generation-rates} already
has established that the Dreicer generation rate is largely insensitive
to the choice of $D^{*}$, provided that it is chosen in the stable
range. Additional proof is provided in Table \ref{tab:D*-insensitivity},
where we list the relative Dreicer growth rates for two values of
$D^{*}$ and two temperatures using the Eulerian-$E$ treatment, demonstrating
the insensitivity of this physical figure of merit to $D^{*}$.

\subsubsection{Scalability and performance of algorithm}

Table \ref{tab:scaling} lists the wall-clock time (WCT) per time
step for the hot (10 keV) plasma at $r/a=0.4$ case with the Eulerian-$E$
treatment, using both global and local interpolation methods in the
orbit integration. The hot case is representative to assess the impact
of the Lagrangian step on performance, as $\epsilon\ll1$ and orbits
are long. The local interpolation scheme is about twice faster than
the global scheme for all grids and processor counts considered, confirming
earlier 0D-2P observations. The Table suggests that the local interpolation
scheme will outperform the global scheme by a larger factor as the
mesh is refined, which is expected. The WCT scales linearly with the
number of degrees of freedom and inversely proportional to the processor
count, demonstrating an optimal, scalable semi-Lagrangian algorithm.
This is expected, as both the Eulerian and the Lagrangian steps are
themselves scalable and optimal. The excellent parallel and algorithmic
scalability of the Eulerian implicit collisional solver has been documented
in Ref. \citep{daniel2020-irfp}. The Lagrangian step is also expected
to be optimal as it integrates one orbit per Eulerian mesh point,
and the cost per orbit scales as $\Delta t\sqrt{D^{*}}/\epsilon$,
independently of mesh refinement. The results in the Table also demonstrate
that our implementation exhibits strong parallel scalability.

\begin{table}
\centering{}\caption{\label{tab:scaling}Total wall-clock time (WCT, in seconds per $\Delta t)$
of the semi-Lagrangian algorithm for the hot (10 keV) case at $r/a=0.4$
using the Eulerian-$E$ treatment with both local and global interpolation
strategies.}
\begin{tabular}{|c|c|c|c|c|c|}
\hline 
Grid ($N_{\theta}\times N_{p_{\parallel}}\times N_{p_{\perp}}$) & MPI tasks & Grid per MPI task & WCT local int. & WCT global int. & WCT ratio\tabularnewline
\hline 
\hline 
\multicolumn{6}{|c|}{Spatial grid-refinement study}\tabularnewline
\hline 
$8\times256\times128$ & 512 & $2\times16\times16$ & 2.7 & 4.8 & 1.77\tabularnewline
\hline 
$16\times256\times128$ & 512 & $4\times16\times16$ & 4.9 & 9.5 & 1.93\tabularnewline
\hline 
$32\times256\times128$ & 512 & $8\times16\times16$ & 9.7 & 18.7 & 1.93\tabularnewline
\hline 
\multicolumn{6}{|c|}{Momentum space grid-refinement study}\tabularnewline
\hline 
$32\times128\times64$ & 512 & $2\times16\times16$ & 2.3 & 4.4 & 1.91\tabularnewline
\hline 
$32\times256\times128$ & 512 & $8\times16\times16$ & 9.7 & 18.7 & 1.93\tabularnewline
\hline 
$32\times512\times256$ & 512 & $32\times16\times16$ & 44.0 & 88.6 & 2.01\tabularnewline
\hline 
\multicolumn{6}{|c|}{Strong parallel scalability study}\tabularnewline
\hline 
$32\times256\times128$ & 128 & $8\times32\times32$ & 34.5 & 67.3 & 1.95\tabularnewline
\hline 
$32\times256\times128$ & 256 & $8\times32\times16$ & 17.7 & 34.3 & 1.94\tabularnewline
\hline 
$32\times256\times128$ & 512 & $8\times16\times16$ & 9.7 & 18.7 & 1.93\tabularnewline
\hline 
\end{tabular}
\end{table}

\section{\label{sec:Conclusions}Conclusions}

We have proposed an asymptotic-preserving, uniformly convergent semi-Lagrangian
numerical scheme for the 2D-2P relativistic collisional drift-kinetic
equation (rDKE). The approach is based on a Green's function reformulation
of the hyperbolic transport component, which leads to a practical
numerical scheme after suitable approximations of quantifiable accuracy
impact. The scheme employs operator splitting to deal with collisional
sources (and perhaps others), which avoids iteration of the stiff
hyperbolic component without spoiling the asymptotic properties. The
collisional step considers the full complexity of relativistic Fokker-Planck
collisions and reuses the scalable and accurate strategy proposed
in Ref. \citep{daniel2020-irfp}. It can be run either in linearized
or in nonlinear mode. The transport step consists of orbit integrals
(lending the approach the semi-Lagrangian character), one per mesh
point, using analytically derived kernels with given fields. Since
one orbit is spawned per Eulerian mesh point, the cost of the Lagrangian
step scales linearly with the number of unknowns, rendering the overall
semi-Lagrangian algorithm both scalable and optimal. By providing
targeted dissipation in the Lagrangian kernels, the method is self-conditioning,
automatically projecting the solution to the slow asymptotic manifold
when the asymptotic regime warrants it. For $\epsilon\ll1$, the method
gives a consistent solution to the limit problem, the so-called bounced-averaged
Fokker-Planck equation. Unlike earlier studies for stiff hyperbolic
transport \citep{crouseilles2013asymptotic,fedele2019asymptotic},
our approach does not require either dimension or variable augmentation,
or iteration in the stiff hyperbolic component. We demonstrate by
analysis and numerical examples in realistic geometries that the scheme
is first-order accurate in $\Delta t$ for $\Delta t<\epsilon$, and
second-order accurate in $\epsilon$ for $\Delta t>\epsilon$. We
have successfully verified the approach against published results
on Dreicer runaway electron generation in axisymmetric cylindrical
tokamak geometries for various plasmas temperatures \citep{mcdevitt2019runaway},
spanning all regimes of interest in $\epsilon$ and demonstrating
the capabilities of the scheme.

While our implementation so far is limited to 2D configurations with
nested flux surfaces, the approach itself features no such limitation,
opening the possibility of the multiscale simulation of runaway electron
dynamics in arbitrary 3D magnetic field topologies with full collisional
physics. Future work will explore such an extension, as well as the
inclusion of additional physics such as radiation forces, knock-on
collisions, etc.

\section*{Acknowledgements}

The authors thank X. Tang and C. McDevitt for help in verifying the
algorithm and other useful insights during the course of this project.
This research used resources provided by the Los Alamos National Laboratory
Institutional Computing Program and is supported by the SciDAC project
on runaway electrons (SCREAM) through the Office of Fusion Energy
Sciences, U.S. Department of Energy. Los Alamos National Laboratory
is operated by Triad National Security, LLC, for the National Nuclear
Security Administration of U.S. Department of Energy (Contract No.
89233218CNA000001).

\appendix

\section{\label{sec:Bounce-averaging-of-rDKE}Relativistic bounce-averaging
operator}

In the relativistic case, the Jacobian of the transformation from
($p_{\parallel},p_{\perp}$) to ($\mathcal{E}=mc^{2}\gamma-e\Phi,\mu$)
is:
\[
\mathcal{J}=\det\left[\frac{\partial(\mathcal{E},\mu)}{\partial(p_{\parallel},p_{\perp})}\right]=\frac{p_{\parallel}p_{\perp}}{m^{2}\gamma B(\mathbf{x})}.
\]
Therefore, by the transformation of probability, $fp_{\perp}dp_{\perp}dp_{\parallel}=gd\mathcal{E}d\mu$,
we have:
\[
g(\mathcal{E},\mu)=\frac{m^{2}B\gamma}{p_{\parallel}}f(\mathbf{x},p_{\parallel},p_{\perp}).
\]
Noting that, along a ($\mathcal{E},\mu$) orbit, the arc-length obeys:
\[
ds=\frac{p_{\parallel}}{\gamma m}d\tau,
\]
we find:
\[
g(\mathcal{E},\mu)\left.\frac{ds}{B}\right|_{\mathcal{E},\mu}=mf\left.d\tau\right|_{\mathcal{E},\mu}\Rightarrow g(\mathcal{E},\mu)=\left\langle f\right\rangle _{\mathcal{E},\mu}=m\frac{\oint_{\mathcal{E},\mu}fd\tau}{\oint_{\mathcal{E},\mu}\frac{ds}{B}},
\]
which defines the relativistic bounce-averaging operator sought.

\section{\label{sec:Local-spline-interpolation}Local spline interpolation
procedure}

To perform second-order local interpolation at a point $(x,y,z)$
in our (non-uniform) logical mesh, we first identify the surrounding
27-point ($3\times3\times3)$ stencil $(x_{i},y_{j},z_{k})$ that
encapsulates said phase-space location, and express the interpolated
value sought as:

\[
p=\sum_{i=0}^{2}\sum_{j=0}^{2}\sum_{k=0}^{2}N_{i,j,k}(x,y,z)p_{i,j,k},
\]
where $N_{i,j,k}(x,y,z)$ is the shape function centered at the node
$(x_{i},y_{j},z_{k})$ (with specific properties, as explained below),
and $p_{i,j,k}$ are the nodal values of the function we wish to interpolate.
Shape functions have unit value at the node in which they are centered
and vanish at all other nodes. Also, they form a partition of unity
(the sum of all shape functions at any interpolated point $(x,y,z)$
is equal to one). For example, the shape function at $(x_{2},y_{2},z_{2})$
is given by:

\[
N_{2,2,2}(x,y,z)=\frac{{(x-x_{1})(x-x_{0})}(y-y_{1})(y-y_{0})(z-z_{1})(z-z_{0})}{{(x_{2}-x_{1})(x_{2}-x_{0})}(y_{2}-y_{1})(y_{2}-y_{0})(z_{2}-z_{1})(z_{2}-z_{0})},
\]
such that

\[
N_{2,2,2}(x_{2},y_{2},z_{2})=1,
\]
and

\[
N_{2,2,2}(x_{i},y_{j},z_{k})=0,\qquad\mathrm{{for}\quad}i,j,k\neq2.
\]
Also,

\[
\sum_{i=0}^{2}\sum_{j=0}^{2}\sum_{k=0}^{2}N_{i,j,k}(x,y,z)=1
\]
for arbitrary $(x,y,z)$.

\pagebreak\bibliographystyle{ieeetr}
\bibliography{iRFP-coll}

\end{document}